\documentclass[prc,aps,final,twocolumn,nofootinbib]{revtex4-2}
\usepackage{graphicx}
\usepackage{xcolor}
\usepackage{amsmath}
\usepackage{lipsum}
\usepackage[utf8]{inputenc}
\usepackage[ukrainian,english]{babel} 
\usepackage[T2A]{fontenc} 
\usepackage[utf8]{inputenc} 
\usepackage[ukrainian,english]{babel} 
\def\be{\begin{equation}}
\def\ee{\end{equation}}
\def\bea{\begin{eqnarray}}
\def\eea{\end{eqnarray}}
\def\l{\label}

\def\cos{\hbox{cos}}
\def\sin{\hbox{sin}}

\def\exp{\hbox{exp}}

\def\d{\hbox{d}}
\def\eps{\varepsilon}
\def\epsi{\mathcal{E}}

\def\epsu{\mathcal{U}}

\def\mitf{\mathcal{T}}

\def\siml{\hbox{\kern.1em \lower.6ex \hbox{$\sim$} \kern-1.12em
 \raise.6ex \hbox{$<$} \kern.1em}}
\def\simg{\hbox{\kern.1em \lower.6ex \hbox{$\sim$} \kern-1.12em
    \raise.6ex \hbox{$>$} \kern.1em}}
\def\mitf{\mathcal{T}}

\begin{document}

\title{MACROSCOPIC
 ROTATIONAL CONSTRAINTS ON THE NEUTRON-STAR RADIUS}   
       
       \author{A.G. Magner}\email{alexander.magner66@gmail.com}
     \affiliation{Nuclear Theory Department,
	Institute for Nuclear Research,
	Prospekt Nauki, 47, Kyiv, 03028, Ukraine}
     
       \author{A.A. Uleiev}
       \affiliation{Nuclear Theory Department,
	Institute for Nuclear Research,
	Prospekt Nauki, 47, Kyiv, 03028, Ukraine}

     \author{S.N. Fedotkin}
     \affiliation{Nuclear Theory Department,
	Institute for Nuclear Research,
	Prospekt Nauki, 47, Kyiv, 03028, Ukraine} 

     \author{S.P. Maydanyuk}
     \affiliation{Nuclear Processes Department,
	Institute for Nuclear Research,
	Prospekt Nauki, 47, Kyiv, 03028, Ukraine}
    \affiliation{Southern Center for Nuclear-Science Theory (SCNT),
	Institute of Modern Physics,
	Chinese Academy of Sciences,
	No. 1 Xinqiao North Road,
	Huizhou 516000, Guangdong Province, China}

    \author{A. Bonasera}
   \affiliation{Cyclotron Institute, Texas A\&M University,
	Luedecke Building, 120 Spence St,
	College Station, TX 77843, USA}

   \author{H. Zheng}
    \affiliation{School of Physics and Information Technology,
	Shaanxi Normal University,
	No. 620 West Chang'an Street, Chang'an District,
	Xi'an 710119, Shaanxi Province, China}

    \author{A.I. Levon}
    \affiliation{Heavy Ions Physics Department,
	Institute for Nuclear Research,
	Prospekt Nauki, 47, Kyiv, 03028, Ukraine}

    \author{U.V. Grygoriev}
  \affiliation{Nuclear Theory Department,
	Institute for Nuclear Research,
	Prospekt Nauki, 47, Kyiv, 03028, Ukraine}

  \author{T. Depastas}
 \affiliation{Cyclotron Institute, Texas A\&M University,
	Luedecke Building, 120 Spence St,
	College Station, TX 77843, USA}

\date{\today}

\begin{abstract}
The macroscopic model for a neutron star
  (NS) as 
 a finite perfect fluid at
  the equilibrium 
  is extended to rotating systems
      by incorporating the linear perturbation expansion 
  over a small frequency $\omega$
  near Schwarzschild outer-inner gravitational metric within the  
  effective-surface (ES)
   approach. 
      The NS angular momentum $I$
      and moment of inertia (MI) 
      for a slow stationary
       azimuthal rotation around 
       the symmetry axis are calculated by using 
       the Kerr metric approach in
            spherical coordinates, and compared with
            Boyer-Lindquist (outer) and Hogan
               (inner) metric results.
               The volume and gradient-surface terms of the
               macroscopic NS energy density 
   $\mathcal{E}(\rho)$ (Equation of State)
   are taken into account 
   at the leading order of  the leptodermic parameter $a/R \ll 1$,
   where $a$ is the ES crust thickness and
  $R$ is the NS effective radius.
         The analytical macroscopic NS MI expressions,
     $\Theta=\d I/\d \omega=\tilde{\Theta}/(1-\mathcal{T}_{t\varphi})$, 
     have been obtained in terms of the
     statistically averaged MI, $\tilde{\Theta}$,
     and its time and 
     azimuthal-angle
     $t,\varphi$  correlation, 
  $\mathcal{T}_{t\varphi}$, as
   sums of the 
    volume and surface 
   components.
    The 
  MI $\Theta$ is changed significantly  
  as function of the effective radius $R$ because
   of  a strong gravity. We found the 
 additional constraint for the NS radius to 
   smaller 
   accessible ranges which is due mainly to  
     the
     $t,\varphi$  correlations and surface contributions.
     The adiabaticity conditions 
         for applicability of the linear
     perturbation theory is carried out 
     for 
    several neutron stars 
    with a strong gravity and relatively large rotation periods.

    \indent
      {\it keywords:} neutron stars, rotation, surface deformations,
      gravitational metric, extended Thomas-Fermi approach,
      moments of inertia
\end{abstract}

\maketitle 

\section{Introduction}

\label{introd}

The macroscopic effective-surface approach
\cite{strtyap,strmagbr,strmagden,MS09,BM13,BM15} (ESA)
taking into account the gradient terms of the
energy density within the Extended Thomas-Fermi approximation
    (ETF) at first order over a small edge thickness $a$ in units of the
    system radius
    $R$ 
    has already been applied \cite{MM24,MM25npa}
for the description of rotating neutron
stars (NSs); see Refs.~\cite{MM26npa,AM25npae}.
Following the original 
  works \cite{RT39,OV39} of Tolman,
Oppenheimer, and Volkoff (TOV) 
we consider the neutron star as
a finite, dense and perfect fluid system at
equilibrium under the gravitational,
and other realistic  forces;
see also the Tolman book \cite{RT87} (chapt. 7, sect. 96).
As shown in Ref.~\cite{RT87},
the TOV equations were derived \cite{OV39} from the
    equations of the
General
Relativity Theory (GRT) assuming the spherical symmetry and the
macroscopic fluid approximation to the energy-momentum tensor $T_\mu^\nu$.
The expression for the
tensor $T_\mu^\nu$ was essentially used in the
TOV derivations  for 
    a macroscopic finite system as
the perfect
fluid system in terms of the energy density functional $\epsi(\rho)$ 
of the mass density $\rho$, and therefore, its gradients
[see Refs.~\cite{LLv2} (chapts. 11 and 12),
\cite{LLv6} (chapt.~1, sects.~1-3; chapt.~7, sect.~61;
chapt.~15, sect.~133) and Rowlinson-Widow \cite{RW82} (chapt.~1)].
Many applications  of the TOV result
    in terms of the outer-inner Schwarzschild metric 
    solutions,  Refs.~\cite{KS16out,KS16in},
for spherically symmetric NSs  can be
found in
Refs.~\cite{LLv2,BBP71,BPS71,WF88,CB97,AA98,PH00Poland,LP01,HP01PRL,HP01PRC,ST04,SH06,HPY07,Ko08,CH08,ZB11,FCPG13,PFCPG13,SG13,RW14,BZ14,BC15,GB18,LH19,KM21prc,KM21aj,GN23,SBL23,Pe23,FG23,DFG23,LJ23,SS23,Pe24,CM25,XV24}
(sect.~9 in Ref.~\cite{ST04} and sect.~3 in Ref.~\cite{HPY07}).
According to numerous observable data, 
the NS mass $M$ is larger than or of the order
of Solar mass but radius $R$ is much smaller 
such that the inner averaged density $\rho$ being
larger than or of the order of that of the nuclear 
    matter; see
Refs.~\cite{TR19,TR21,MM19,AD24}. 
This might be a statistical
background of the TOV assumption that the
NS is similar
to the finite perfect dense fluid  similar to nuclear or 
molecular water (amorphous solid) systems \cite{BBP71,BPS71}.
     For a good hint to the nuclear astrophysics; 
     see also, e.g., many observable data
     \cite{DK15,JN17,TMT17,BA19,BA19-1,GR19,CC20,MM21,VD22,RK23,DS24,MB13,YK24}
     and theoretical 
   \cite{BPS71,PH00Poland,LP01,HP01PRL,HP01PRC,FCPG13,PFCPG13,SG13,RW14,BZ14,BC15,GB18,LH19,KM21prc,KM21aj,GN23,SBL23,Pe23,FG23,DFG23,LJ23,SS23,Pe24,XV24} works.
   Clear specific definitions and complete updated results
    for the energy density $\epsi(\rho)$ with
density gradient (surface) terms and  for equations of
the infinite matter state
 with many inter-particle
forces in the non-relativistic and relativistic cases
can be found in the 
recent reviews Refs.~\cite{GB18,SBL23}.

Using the step-function for the density $\rho$, 
    which is a
    constant inside and zero outside of the system, in line of the
    Schwarzschild
    derivations \cite{KS16out,KS16in},
 Tolman obtained \cite{RT87}
 the explicit analytical solutions for the pressure.
    They were found in terms of the
 Schwarzschild metric
 coefficients for the spherical system symmetry, and the
 macroscopic energy-momentum tensor $T_\mu^\nu$ for
 the perfect fluid.
 Analytical expressions for the pressure in terms of the
 simple one-dimensional integrals can be obtained
 also for the continuous leptodermic
 density $\rho$. 
 Notice that the upper Schwarzschild radius\footnote{
         This is in
     contrast to another Schwarzschild gravitational radius, $r_g\propto 2M$,
     which is the lowest constraint for the NS radius, $r_g < R < R_{\rm S}$.}
     (see Refs.~\cite{RT87,MM24,MM26npa})
 $R_S\propto (\epsi(\overline{\rho}))^{-1/2}$,
    can be considered approximately as
    a constant which determines the upper limit of the NS radius,
        $R<R_{\rm S}$,
           if the density $\rho$
    has a saturation property $\rho \approx \overline{\rho}$ inside of the
     stable NS. Therefore, the  energy density $\epsi$ which is
    the argument of $R_S$
       should be assumed \cite{RT87} to be
    also the corresponding saturation property.

      As the TOV equations \cite{OV39} are used along with
    the Equation of State (EoS) in a lot
    of astrophysical works on the NS,
    it would be logical to
    agree the arguments for the
    specific derivations of both
    these equations. This is the main motivation of our ESA and TOV
    macroscopic approaches \cite{MM24,MM25npa,MM26npa,AM25npae}. 
    Notice also that, of course, in nuclear astrophysics 
    we have to take into account a
    strong gravitational field as in 
        the derivations
    of the TOV equations,  unlike
        the traditional
    nuclear physics. Our suggestion 
    for this coordination
    does not exclude more microscopic EoSs. However, it requires 
    the corresponding
    modification of the TOV equations 
        which, unfortunately, makes the problem
        to be more complicate than 
        the
     present analytical approach. 

     The gradient surface 
 components of the energy density $\epsi(\rho)$ 
 were suggested basically in many works; see, e.g.,
 Refs.~\cite{BBP71,BPS71,LP01,HP01PRL,HP01PRC,PH00Poland,FG01,BB03,HPY07,CH08,ABCG09,ZB11,PFCPG13,FCPG13,SG13,BZ14,FG23,DFG23,SBL23,LJ23,Pe23,Pe24,XV24,MM24}.        
 Within the leptodermic  approximation \cite{MM24,MM25npa,MM26npa} based on such
     gradient density components of the $\epsi(\rho)$, 
         one can consider
     first analytically a dense one-component system. 
     In these calculations we take the energy 
     density functional $\epsi(\rho)$
 in a rather
 general form,  as linear in
     $(\nabla \rho)^2$ terms, with the coefficients
 which are smooth functions
 of the density $\rho$.
 In such a system, the NS effective surface (ES) can be defined as
 the inflection points of the maximum of $(\nabla \rho)^2$. The
 density
 $\rho$ in the leading order
 approximation is function of the
 perpendicular-to-ES coordinate with
exponentially decreasing behavior 
from almost a constant saturation value inside of the dense system
to that of zero continuously through the ES in
a relatively small crust range $a$.
In order to obtain the analytical solutions for the
density and EoS, we 
use the
 leptodermic macroscopic
 approximation.
 In this approximation,
the gravitational field
can be taken into account through the gravitational energy density 
part $\mathcal{U}(\rho)$, determined by 
    the second order expansion over deflection
of the density $\rho$ from its saturation value
$\overline{\rho}$
(see 
Refs. \cite{MM24,MM25npa,MM26npa}).
Using the approximation $a/R \ll 1$,
simple and
accurate solutions of many nuclear
\cite{strtyap,strmagbr,strmagden,MS09,BM13,BM15},
molecular (amorphous solids) \cite{RW82,vdW},
and metallic cluster \cite{FG01,BB03}
problems 
were obtained by using the ESA; 
see also Refs.~\cite{MM24,MM25npa,MM26npa} for applications
to the neutron stars.
The ESA exploits
the property of saturation of the 
continuous density $\rho$
inside of the system,
which is a characteristic
feature of any such dense systems.
As a result, 
relatively narrow edge region exists in these systems
in which the 
density drops continuously but relatively much
from its almost central value to zero,
while the part 
inside of the system far from the ES can be  changed
a little (saturation property of a dense system).

The condition of the NS
variational 
    equilibrium
      means that 
      the variation of
      the total energy $E$ over the
density $\rho$ is zero under the constraints which fix 
some
integrals of
motion beyond the energy $E$ by the Lagrange method. The Lagrange multipliers
  can be determined by these constraints
  within the local energy-density theory, in particular,
  the extended Thomas-Fermi 
  (ETF)  approach 
   \cite{brguehak,BB03}.
The variational equation mainly determines approximately the density
distribution 
across the diffused surface layer. The latter is of the relatively
small order over
the ratio of the diffuseness 
    parameter $a$ 
to the (mean curvature) ES radius $R$,
$a/R << 1$, for sufficiently
heavy systems. 
A small parameter, $a/R$, of the expansion 
within the ESA 
can be used
for analytical solving the variational problem
for the minimum of the energy density 
with constraints for
a fixed particle number, and other integrals of motion,
 such as angular momentum,
 quadrupole deformation, tidal deformations of the binary NSs,
 etc.   
When this edge distribution of the 
density is 
derived, the leading
static and dynamic density distributions which 
correspond to the
diffused surface conditions can be easily constructed. 
 Our ESA
is
based on the catastrophe theory for solving differential equations with
a small parameter of the order of $a/R$ as the coefficient
in front of the high order derivatives in normal to the ES direction.
     A relatively large change of the density $\rho$ on a small
     distance $a$ with respect to the curvature 
     radius $R$ takes place for
      finite dense
 liquid-matter systems mentioned above.
 Inside of such dense systems,
the density $\rho$ is changed a
little around
constant saturation density $\overline{\rho}$. Therefore, including also
the second order expansion of the gravitational
energy density
component $\mathcal{U}(\rho)$ near $\overline{\rho}$,
one obtains
essential effects
of the surface capillary pressure 
    which were originally studied for dense
molecular systems by van der Waals; see
Refs.~\cite{RW82,vdW}.

The accuracy of the ESA was checked in 
Ref.~\cite{strmagden} for the case of 
    static and dynamic nuclear physics
by comparing the results 
with the existing nuclear theories like Hartree-Fock (HF)
\cite{vauthbrink} and ETF \cite{brguehak,BB03}, based on the 
Skyrme forces 
\cite{vauthbrink,skyrme,barjac,ringshuk,blaizot,brguehak,gramvoros80,krivin,CB97,CB98},
but for the simplest
case without spin-orbit and 
asymmetry terms of the energy density functional.
The extension of the ES approach
to the nuclear isotopic asymmetry and spin-orbit
 interaction has been done in
 Refs. \cite{MS09,BM13,BM15}. The Swiatecki derivative
terms of the symmetry energy for
heavy nuclei
\cite{danielewicz2,vinas14}
were 
taken into account  within the ESA in Ref. \cite{BM15}. The
discussions of 
a progress in nuclear physics and astrophysics
within the relativistic local density approach,
can be found in review articles,
Refs.~\cite{GB18,SBL23,NVR11}; see also
Refs.~\cite{CH08,Pi21,Pe23,Pe24}.  
 In Ref.~\cite{MM24}, 
within the effective surface approach
for the neutron stars, 
the leptodermic corrections to TOV equations were derived.

 The macroscopic NS rotation, as a small 
 rotational-energy perturbation
 of the strongly gravitating spherically-symmetric 
 background described 
 by the Schwarzschild metric \cite{KS16out,KS16in}
     was considered \cite{AM25npae,MM26npa} within
 the Kerr approach  
 \cite{RK63}
  for a relatively 
        slow NS rotation.            
   More simple coordianates were suggested by Boyer and Lindquist
        \cite{BL67} for the region outside of the 
         NS ($r>R$), and by Hogan
        for the inner NS part \cite{PH76,PC78} ($r<R$); see
        also Refs.~\cite{LLv2,Teu15}.  
            Independently, the  
        specific perturbation  formulation based on the non-rotating
        Schwarzschild gravitational metric and taking into  account
        small quadrupole- (or spheroidal-) type
        surface deformations
        due to a  small uniform
        frequency-fluid  rotation
        was suggested by Hartle and Thorne in
        Refs.~\cite{TC67,JH67,HT68} based
        on Regge and Wheeler Ref.~\cite{TR57};
        see also Ref.~\cite{AK78}. The 
        Friedman, Ipser and 
        Stergioulas formalism
        for the description of the
    uniformly rotating neutron stars
    with a constant angular velocity $\omega$ 
    was presented for a small angular momentum of
    the perfect stellar fluid
    \cite{FIP86,NS03,AW08,NS17}.

    In the present work we derive the coordinate dependence of
    a Kerr component of the gravitational
    metric $g^{}_{03}(r,\theta)$
    from the GRT
    in the linear approximation over
        the angular momentum $I$, 
        and
    the corresponding moments of inertia.
The general macroscopic theory for the NS
    rotations based on the Kerr approach is presented in Sect.~\ref{KGRT}.
     The GRT solutions
    for the rotational component of gravitational axially-symmetric
    metric $g^{}_{03}(r,\theta)$
    are derived  in the simplest
    linear approximation over
     a small angular momentum parameter $I/M^2$ (Sect.~\ref{LR}). 
    The results for the 
        moment of inertia (MI) calculations for the deformed NS
    in terms of the volume and surface components are obtained in
    Sect.~\ref{SFA}.
    They are discussed in
    Sect.~\ref{disc} and
 summarized in
 the concluding
 section \ref{concl}. Some details of the calculations
are given in
Appendices \ref{appA}-\ref{appD}.

\section{Kerr GRT approach for a rotating NS}
\l{KGRT}

\subsection{
General points for macroscopic gravitational fluid systems}
\l{GRT}

For the derivations of the gravitational metric $g_{\mu\nu}$
we may begin with the
simplest GRT equation
\cite{RT87,LLv2}:
\be\l{GRTeq}
-\frac{8 \pi G}{c^4} T_{\mu}^{~\nu}=\mathcal{R}_{\mu}^{~\nu} 
-\frac{1}{2}g_{\mu}^{~\nu}Tr\mathcal{R}~,
\ee
where $\mathcal{R}_{\mu}^{~\nu}$ is the
standard tensor Ricci, 
$Tr\mathcal{R}=g_{\mu\nu}\mathcal{R}^{\mu\nu}$, $G$ is the
gravitational constant, and $c$ is the speed of light.
The energy-momentum tensor for the perfect dense liquid 
drop, $T_{\mu}^{~\nu}$, is given by \cite{LLv6}
\be\l{Tmn}
T_{\mu}^{~\nu}=\mathcal{E}(\rho)u_\mu u^\nu 
+\mathcal{P} g_{\mu}^{~\nu},
\ee
where $\mathcal{P}$ is the pressure, 
$\epsi(\rho)$ 
is the energy density
in the free energy ensemble, and $u_\mu$ the particle
 four-velocity.
 For the linear perturbation theory (LPT) calculations of the
moment of inertia (MI)
we need the macroscopic energy density $\epsi(\rho)$ at zero
frequency $\omega$,
where $\rho$ is the mass density, and
    $u_4\sim \omega$ in spherical coordinates.
We will use a general macroscopic expression
for the energy density of a fluid system \cite{LLv6} (see also
    Appendix \ref{appA}): 
    \begin{equation}
      \label{enerden}
\epsi\left(\rho\right) =\mathcal{A}(\rho)
+\mathcal{B}(\rho)\left(\nabla \rho\right)^2~,
\end{equation}
where $\mathcal{A}(\rho)$
and $\mathcal{B}(\rho)$ are smooth 
functions of the density $\rho$. 
A non-gradient part $\mathcal{A}(\rho)$ of the energy density
$\epsi$ 
can be presented (at $\omega=0$) as that done in
Refs.~\cite{MM24,MM25npa,MM26npa}.
Within our macroscopic leptodermic approach, one can introduce 
the total energy density  $\epsi(\rho)$, including its gravitational
part \cite{MM24,MM25npa,AM25npae,MM26npa}
$\mathcal{U}(\rho)$.
Then,
    for leptodermic systems, one can expand~~$\epsu(\rho)$
over powers of
the difference $\rho-\overline{\rho}$
   near
the mean 
density
$\overline{\rho}$, inside of the ES, up to 
second order,
\be\l{phiexp}
\epsu(\rho)=
\epsu_0+\frac12~\epsu_2(\rho-\overline{\rho})^2~,
\ee
where 
$\epsu_0=\epsu(\overline{\rho})$
and~~
$\epsu_2=
\partial^2 \epsu(\overline{\rho})/\partial \rho^2$
are derivatives of the~~$\epsu(\rho)$ over $\rho$ 
    taken at the
    inner density $\overline{\rho}$.
Linear terms
disappear because of 
the condition of
minimum of the energy per particle for
an isolated system; see, e.g., Ref.~\cite{MM24}.
 With the 
expansion, Eq.~(\ref{phiexp}),
for $\mathcal{A}(\rho)$
one obtains   
\be\l{vol}
\mathcal{A}=\rho c^2 
+\eps^{}_G(\rho)~,
\ee
where     $\eps^{}_G(\rho)$ is the total second-order
energy density 
    component, 
modified
by the gravitational field,  
\be\l{epsKG}
\eps^{}_G(\rho)=\eps(\rho)+\frac{1}{2}
\epsu_2 \rho(\rho-\overline{\rho})^2
=\frac{K_G}{18m \overline{\rho}^2}~
\rho\left(\rho-\overline{\rho}\right)^2~. 
\ee
Here, $\eps(\rho)$ is given by  
\be\l{eps}
\eps(\rho)=\frac{K}{18 m \overline{\rho}^2}~
\rho\left(\rho-\overline{\rho}\right)^2~,
\ee
where $\overline{\rho}=m \overline{n}$, $m$ is the test particle mass,
$\overline{n}$ is the 
    statistically averaged particle-number density
$n$ inside of the system far from the ES,
$K_G$ is the total incompressibility modulus modified by the
gravitational field ($K_G>0$),
\be\l{KG}
K_G=K+9 \overline{\rho}^2\epsu_2~,
\ee
and $K$ the non-gravitational (e.g., nuclear) 
incompressibility
    component.
     Notice that the 
    energy density $\epsi(\rho)$, Eq.~(\ref{enerden}),
     is 
    general, in
    particular,
    for all known Skyrme forces in nuclear physics taking
    into account the volume and surface terms.
    Another famous example is
    the 
        van der Waals forces for a dense 
        molecular system \cite{RW82} (e.g., the
        liquid drop).    
    The coefficient $\mathcal{B}(\rho)$ in front of
the gradient squared term
in Eq.~(\ref{enerden}) is given by
\be\l{surf}
\mathcal{B}(\rho) = \mathcal{C}+\mathcal{D}\rho +
\frac{\Gamma}{\rho} \approx \mathcal{C}~.
\ee
These
terms are associated with the nuclear Skyrme interaction modified by
the  gravitational component. First term is related to the
interaction term which is a main reason of
the diﬀuse surface thickness
for a non-rotating NS. In this sense it has a more general meaning
including the
main eﬀective interaction in a dense molecular system, which was
studied by van der
Waals \cite{vdW,RW82}.
Therefore, we will call this component as that of
the vdW-Skyrme interaction. The second term is coming from the spin-orbit
interaction. This interaction might be important, for instance, for
any dense rotating
liquid-drop systems with a sharp surface edge, i.e., with large
density-gradient terms
in the surface region for a ﬁnite leptodermic systems; see also the
same general arguments for using the ETF approach in
Refs.~\cite{BB03,brguehak}.
The last term is a gradient
correction to the kinetic energy ($\hbar^2$ correction of the 
kinetic energy in the general
ETF approach \cite{BB03,brguehak}); see the Wilets result for the 
density $\rho$ in 
Refs.~\cite{MM24,MM26npa}.
Thus, the forms Eq.~(\ref{vol}) 
for $\mathcal{A}$ and Eq.~(\ref{surf}) 
for $\mathcal{B}$ are rather general among the
simplest analytical solutions for any dense ﬁnite systems.
    (In Ref.~\cite{MM24}, the inter-particle interaction
    constant was defined as $m^2\mathcal{C}$ because $\rho$
    is there the particle-number density.)
The constant $\mathcal{C}$, associated with the
nuclear Skyrme and molecular van der
Waals (vdW-Skyrme) interaction \cite{brguehak,BB03}, is related mainly to the
diffuse NS surface thickness,
$a\propto (\mathcal{C}\overline{\rho}/m K_G)^{1/2}$
(from Refs.~\cite{MM24,MM26npa}), see Eq.~(\ref{leppar}).
For simplicity, one can neglect the 
correction corresponding to
the spin-orbit term known well from 
the nuclear physics; see e.g., 
Refs.~\cite{ringshuk,MM24}.
We neglect here also
    a relatively small  surface gradient correction
    to the kinetic energy which is important for 
    inner gas  system \cite{MM24,MM26npa}
    which essentially differs
from our desired dense fluid one.
It is associated with the
nuclear Skyrme and molecular van der
Waals (vdW-Skyrme) local interaction \cite{brguehak,BB03},
under a strong gravity. 

We should add the constraint for the variational procedure to
get equations
for the static equilibrium. For non-rotated one-component
NS-system example, one should have
to fix the total particle number $N$.
Introducing the chemical  potential,
$\mu=\mu^{}_0$, as the Lagrange
multiplier function of the total particle number, 
    one obtains from Eqs.~(\ref{energytot}) and (\ref{enerden})
    the equation
    for the NS equilibrium - EoS at zero rotational
    frequency $\omega=0$,
\be\l{eq}
\hspace{-0.5cm}\frac{\delta \epsi}{\delta \rho}\equiv
\frac{\partial\mathcal{A}}{\partial \rho}
-2 \frac{\mathcal{C}}{m}~\Delta \rho =\mu~,
\ee
where $\mathcal{A}(\rho)$
  is  given by Eq.~(\ref{vol}).
For nuclear
liquid drop, one has two equations related to the two
constraints for
the
fixed
neutron and proton numbers
 (Refs.~\cite{MS09,BM13,BM15}). 
They determine two (neutron and
proton) chemical potentials.
The Coulomb interaction can be taken into account for a proton
part of the
nucleus through the Coulomb potential; see Ref.~\cite{MS09}.
 Equation (\ref{enerden}) for  $\mathcal{E} $
 overlaps most of the vdW-Skyrme forces \cite{vdW,CB97,CB98}.
The symmetry energy, and surface 
terms
 can be also taken into account in nuclear astrophysics
 similarly as in nuclear physics.
We should emphasize also
that there is, in fact, no
separation of the nuclear and gravitational contributions into the
volume
and surface energies
because of the $\rho$ density dependence of the gravitational 
energy density $\epsu(\rho)$, Eq.~(\ref{phiexp}),
and the Lagrange equation (\ref{eq})
 for $\rho$. 
  The smallness factor, proportional to $a/R$, with
  respect to the volume contribution appears
    because of the integration over the normal-to-ES
  (radial in this example) variable $(r-R)/a$ in
  this ES region.  Thus, 
      for the calculations of the surface corrections to the
NS total energy $E$,
Eq.~(\ref{energytot}), and particle number, 
it would seem possible to assume 
that they
are not (very) important 
for
so small relative diffuseness thickness $a/R\sim 0.1$.
However, as shown in Refs.~\cite{MM24,MM25npa,AM25npae,MM26npa},
the influence of the NS surface becomes really important
for the local
  quantities as the density $\rho(r)$, 
  the EoS,
      and therefore, for the pressure $P(r)$ 
         calculating through
         Eq.~(\ref{enerden}); see 
             Refs.~\cite{MM24,MM25npa,MM26npa}.
Moreover, the gravitational mass defect due to the radial curvature
of the nonlinear Schwarzschild metric in calculations of the NS 
    mass $M$
leads to its dramatic non-monotonic 
behavior as function of the effective radius, $M=M(R)$;
see Refs.~\cite{MM24,AM25npae,MM26npa}. 
    and Appendix
\ref{appA}.

Notice that the explicit expressions
for the tensor Ricci $\mathcal{R}_{\mu\nu}$ 
in terms of the 
gravitational metric $g_{\mu\nu}$ through the Cristoffel 
symbols can be found, e.g., in
Ref.~\cite{LLv2}; see also Ref.~\cite{MM26npa}.
Notice also that a more general GRT $f(\mathcal{R})$ approach 
is based on the nonlinear (e.g., quadratic, over the Riemann tensor)
Lagrange equation, Ref.~\cite{VZ24}. This approach might be 
a challenge to consider 
the NS rotations  in a
forthcoming work.  

     \subsection{The outer Kerr metric approach}
        \l{KBL}

     The  metric tensor solutions to the
        GRT equation (\ref{GRTeq}) for the 
         uniform rotating
        NS around the symmetry axis
         at 
        a small rotational frequency $\omega$
        by carefully taking into account a frequency and
        deformation dependence of the 
        gravitational
        field was 
          suggested \cite{RK63,Teu15} by Kerr.   
        By using a transformation to 
        the spherical 4-coordinate system
        $\{t,r,\theta,\varphi\}$,
        the Kerr metric solution 
        \cite{RK63} was presented by
        Boyer and Lindquist
        \cite{BL67} in simpler coordinates
      (see also Refs.~\cite{Teu15,LLv2}). In the linear
        approximation over $\omega$, one has the space-time
        element:
\bea\l{AWmetgen}
&{\rm d} s^2=e^\nu c^2{\rm d} t^2
+2\tau \alpha^{}_K \sin^2\theta c\d t \d \varphi
-e^\lambda{\rm d} r^2\nonumber\\
&-r^2 {\rm d}\theta^2-
     r^2 \sin^2\theta {\rm d} \varphi^2~,
     \eea
where $\nu(r)$, and $\lambda(r)$ are parameters of 
the Schwarzschild metric. 
    With explicit expressions for these parameters,
one finds  
  \bea\l{BLm}
 &\d s^2 =\left(1 -\frac{r_g}{r}\right)
  c^2\d t^2 +2~\alpha^{}_K~\tau\sin^2\theta~c\d t\d \varphi\nonumber\\
& -\frac{1}{1-r_g/r}\d r^2
-r^2 \d \theta^2-r^2\sin^2\theta \d \varphi^2,~~~r>R.
 \eea
 Here, $r_g$ is the gravitational radius,
 \be\l{rg}
r_g=\frac{2MG}{c^2}~,
 \ee
$M$ is the NS mass,
   $\alpha^{}_K$ is Kerr's
  rotational-frequency constant parameter,
  approximately $\alpha^{}_K \propto \omega$,  
 and $\omega$ is the 
 azimuthal rotational frequency.
  According to this approach, one has
 the approximate relation 
  between the parameter $\alpha^{}_K$ 
  and the NS angular momentum $I$,
 \be\l{OmegaK}
 \alpha^{}_K Mc \approx I = \omega\Theta~,
 \ee
 where $\Theta$ is the NS moment of inertia (MI).
 It is convenient to introduce
 the dimensionless rotation-frequency
 (or angular-momentum) parameter \cite{LB11,PP12},
 \be\l{rotpar}
 \overline{\omega}=\frac{\alpha^{}_K c^2}{M G}
 \approx \frac{Ic}{M^2G}=
 ~\frac{\Theta c}{M^2G}~\omega \ll  1~.
 \ee
 We will consider the perturbation approach, linear over
 this rotational parameter.
 In Eq.~(\ref{BLm}),
 $\tau$ is generally
 speaking a dimensionless function of two variables $r$ and $\theta$ for
 an axially-symmetric
 system \cite{RK63}, $\tau=\tau(r,\theta)$; see also the explicit
 equation for $\tau(r,\theta)$ in partial derivatives in
 Ref.~\cite{JH67}, based on Regge and Wheeler vacuum solutions of the
 GRT equation (\ref{GRTeq}). Boyer and Lindquist found
 \cite{BL67} in the spherical vacuum approximation  the
 following 
 expression for  $\tau$ as function of only one variable $r$
 (see also Refs.~\cite{LLv2,Teu15}),
\be\l{tauout}
\tau=r_g/r~.
\ee
One of the main aims of this paper is to derive this solution
more constructively 
    and generally, accounting for the NS surface deformations, 
under the specific
linear approximation over $\overline{\omega}$, Eq.~(\ref{rotpar}),
$\overline{\omega}\ll 1$, outside of the NS for $r>R$ for a finite effective
NS radius $R$,
starting from the linear GRT
equation (\ref{GRTeq}). Another our aim is to solve this GRT equation in the
inner NS part in the same leading approximation,  $\overline{\omega}\ll 1$
(see the next section \ref{DRNSM}).
The metric solution (\ref{BLm}) has the well-known
outer ($r>R$) Schwarzschild metric 
  limit \cite{KS16out,RT87,LLv2} for
 $\overline{\omega} \rightarrow 0$:
     \bea\l{SchwarzOUT}
  & \d s^2
  \rightarrow \left(1 - \frac{r_g}{r}\right)
 c^2\d t^2 -\left(1-\frac{r_g}{r}\right)^{-1}
 \d r^2\nonumber\\
 & - r^2 \left(\d \theta^2 +
  \sin^2 \theta \d \varphi^2\right).
  \eea
For the outer Schwarzschild 
metric element, Eq.~(\ref{SchwarzOUT}),
one has $\nu$ and $\lambda$ at $r>R$ in the leading leptodermic
approximation 
(see Eq.~(\ref{SchwarzOUT}) and Ref.~\cite{RT87}), 
\be\l{nulamout}
\nu_{\rm out}=\log\left(1-\frac{r_g}{r}\right),\quad
\lambda_{\rm out}=-\nu_{\rm out}~. 
\ee
Thus, the gravitational
metric $g^{}_{\mu\nu}$, Eq.~(\ref{BLm}) 
(Ref.~\cite{BL67}),
is the outer Kerr metric for the neutron star with the finite
mean effective radius $R$.

\subsection{The inner Kerr metric approach}
\l{KHM}

For the inner Kerr metric $g^{}_{\mu\nu}$,
going to the inner Schwarzschild metric 
in the limit $\omega \rightarrow 0$ (see Ref.~\cite{RT87}), 
the gravitational metric was suggested by Hogan 
\cite{PH76,PC78}. 
    In the linear approximation over $\omega$, one can present
the space-time element (\ref{AWmetgen}) as
\bea\l{inkerrmet}
&\d s^2=f^{}_{\rm S} c^2\d t^2 +2\tau\alpha^{}_K\sin^2\theta c\d t\d \varphi
-\frac{1}{1-r^2/R^2_{\rm S}}\d r^2\nonumber\\
&-r^2\d \theta^2-r^2\sin^2\theta\d\varphi^2,~~~r<R,
\eea
where
\be\l{fchiQ}
f^{}_{\rm S}=\left(A_{\rm S}-
\frac12\sqrt{1-\frac{r^2}{R^2_{\rm S}}}\right)^2,
\ee
     $A_{\rm S}$ is a constant,
 \be\l{AS}
      A_{\rm S}=\frac32\sqrt{1-\frac{R^2}{R_{\rm S}^2}}~,~~
      \ee
          and $R_{\rm S}$ is the Schwarzschild 
          radius 
              [$R$ in notations of Ref.~\cite{RT87}, in contrast to the
                 gravitational
                radius $r_g$, Eq.~(\ref{rg})],
\be\l{RS}
R_{\rm S}=\sqrt{\frac{3 c^4}{8\pi G \epsi_0}}~.
\ee
    Here, $\epsi_0$ is the internal energy density \cite{MM24,MM25npa},
\be\l{epsi0}
\epsi_0=\epsi(\overline{\rho})
\approx \overline{\rho} c^2~.
\ee

In the nonrotating limit, $\omega \rightarrow 0$, one obtains the inner
Schwarzschild metric 
\cite{RT87} at $r < R$,
     \bea\l{SchwarzIN}
      &  {\rm d} s^2=c^2\left[A_{\rm S} -
       \frac12\sqrt{1- r^2/R^2_{\rm S}}\right]^2 {\rm d} t^2
     -\frac{{\rm d} r^2}{1-r^2/R^2_{\rm S}}\nonumber\\
     &-r^2 {\rm d}\theta^2 -
     r^2 \sin^2\theta {\rm d} \varphi^2~. 
     \eea
 Then, for 
the inner solutions of $\nu$ and $\lambda$ in Eq.~(\ref{AWmetgen}),
one finds
\be\l{nulamin}
\nu_{\rm in}=2\log\left(A_{\rm S}-\frac12\sqrt{1-\frac{r^2}{R_{\rm S}^2}}\right),
~~
\lambda_{\rm in}=-\log\left(1-\frac{r^2}{R_{\rm S}^2}\right).
\ee

\section{NS rotations as a linear response}
\l{LR}
    
\subsection{The NS angular momentum}
\l{AM}
Let us use the LPT
expansion over a
    small parameter
    $\overline{\omega}$;
    see Eq.~(\ref{OmegaK}). 
For the angular momentum $I$ in the case of the stationary rotation
of a perfect liquid drop around its symmetry axis
under a strong gravitational field,
one can write \cite{FIP86,NS03}
\be\l{Igen}
I=\frac{1}{c^2}\int T_\mu^{~\nu}~\phi^\mu 
\hat{n}_\nu  
\d \mathcal{V}_{\omega}~,
\ee
where
$ T_{\mu}^{~\nu}$ is the energy-momentum tensor of the
perfect fluid;
see Eq.~(\ref{Tmn}).
Here, the Killing vector $\phi^\mu$ is defined 
as that acting in
the azimuthal direction reflecting axial
symmetry, and $\hat{n}_\nu$ is the normal vector to the
spatial volume hyper-surface in the 4-space. In Eq.~(\ref{Igen}), the
integral is taken over the whole three-dimesional space.

As shown in Appendix \ref{appB},
 using the LPT approximation, Eq.~(\ref{rotpar}),
 one can neglect all $\overline{\omega}^2$ 
 corrections to the linear 
 terms in the angular momentum $I$, 
Eq.~(\ref{Igen}),  because 
its integrand is in turn approximately proportional to our
small dimensionless
parameter,
$\overline{\omega}$, see Eq.~(\ref{rotpar}). In particular,
the integration in Eq.~(\ref{Igen}) is approximately 
carried out in fact over
the volume element $\d \mathcal{V}$, which is equal to
$\d \mathcal{V}_{\omega}$ at $\omega=0$,
\bea\l{dV}
&\d \mathcal{V}_\omega=
\sqrt{g_{rr}g^{}_{\theta\theta} g_{\varphi\varphi}}
\d r \d \theta \d \varphi\approx
\d \mathcal{V}\nonumber\\
&=e^{\lambda/2} 
\d {\bf r}=J(r) r^2 \d r \sin \theta \d \theta \d \varphi~.
\eea
The radial volume shortness
Jacobian $J(r)$ is determined by the outer or inner Schwarzschild metric, 
Eqs.~(\ref{SchwarzOUT}) and (\ref{SchwarzIN}), respectively:
\bea\l{Jin}
J(r)&=& \frac{1}{\sqrt{1-r^2/R_{\rm S}^2}}~~~(r\le R)\\
&=&\frac{1}{\sqrt{1-r^{}_g/r}}~~~(r > R)~,
\l{Jout}
\eea
  where $r^{}_g$ is the 
 gravitational radius, Eq.~(\ref{rg}), and $R_{\rm S}$ 
 is the Schwarzschild radius, Eq.~(\ref{RS}).
The unit
time-like  four-velocity $u^\mu$ in Eq.~(\ref{Tmn}) 
obeys the normalization condition $u^\mu u_\mu=-1$.
For uniform slow rotations of the axially symmetric
NSs,  it takes the form
$u^\mu = u^t\left(1, 0, 0,\overline{\omega}\right)$,
where $\overline{\omega}$ is given\footnote{In
these derivations, 
the final result
for the MI, Eq.~(\ref{MI}), does not depend on the choice of
a dimension of the frequency $\omega$  
for the considered small
$\overline{\omega}$ because one can include
the constant of the ratio of that to the 
corresponding dimensionless quantity 
into the
definition of the angular momentum $I$ 
itself, Eq.~(\ref{Igen}).}
    by Eq.~(\ref{OmegaK}).
The
gravitational metric element $g_{t\varphi}$ for a small rotational
   parameter $\overline{\omega}$, approximately proportional to the
angular momentum $I$ by
Eq.~(\ref{OmegaK}),
$\alpha^{}_K \propto \overline{\omega} \propto I$, can be
presented similarly
as that in Refs.~\cite{AW08,NS03,NS17};
see Eq.~(\ref{AWmetgen}). 

As shown in Appendix \ref{appB}, after some simple tensor-algebra
transformations, up to $\overline{\omega}^2$ terms,
the NS MI can be calculated as
\be\l{MI}
\Theta=\frac{\partial I}{\partial \omega}=
\frac{\tilde{\Theta}}{1 - \mitf_{t\varphi}}~,
\ee
where $\tilde{\Theta}$ is the statistically 
averaged MI, 
\be\l{MIt}
\tilde{\Theta}=\frac{1}{c^2}
\int  \epsi(\rho)e^{-\nu} r_\perp^2\d \mathcal{V}~,
\ee
and $\epsi(\rho)$ is the energy density, Eq.~(\ref{enerden}),
as functional of the
mass density $\rho$, $\epsi=\epsi(\rho)$ is the NS macroscopic
Equation of State for a non-rotating NS,
 i.e.,
in the limit
$\alpha^{}_K\rightarrow 0$ ($\overline{\omega} \rightarrow 0$). 
In Eq.~(\ref{MI}), $\mitf_{t\varphi}$ is the 
time and azimuthal angle ($t,\varphi$)
correction
due to
the off-diagonal gravitational metric, $g_{t\varphi}$,
\be\l{MItf}
\mitf_{t\varphi}\approx \frac{2}{Mc^2}
\int \epsi(\rho)e^{-\nu} \frac{\tau}{r^2}
r_\perp^2 \d \mathcal{V}~,
   \ee
   where $\tau$ is a function of the radial $r$ and angle
   $\theta$ coordinates in the
   off-diagonal component $g_{t\varphi}$ of the metric $g_{\mu\nu}$;
see, e.g.,
  Eq.~(\ref{tauout}) in the spherical, 
  and derivations below in the deformed case.
  Notice that in Eqs.~(\ref{Igen}), (\ref{MIt}) and (\ref{MItf}) , the
integral is taken over the whole three-dimesional space.
 In Eqs.~(\ref{MIt}) and (\ref{MItf}),
$\nu$ is defined by the Schwarzschild gravitational metric,
Eqs.~(\ref{BLm}) and (\ref{SchwarzIN}), in the form 
(\ref{AWmetgen}) in our linear approximation over
$\overline{\omega}$, and
$r^{}_\perp=r~\sin\theta$. The pole of the MI, Eq.~(\ref{MI}) appears
because of the self-consistency condition (\ref{OmegaK}); see Appendix
\ref{appB}.
This pole is the reason of the additional more narrow
constraints for the NS radius;
see more detailed discussions below.

     Using the statistical averaging over the canonical ensemble
\cite{LLv5}, one
can exclude the off-diagonal correlation term,
$\propto \d t\d \varphi$, from the gravitational metric,
Eq.~(\ref{AWmetgen}). Removing also the term $g_{t\varphi}$ from
parentheses of Eq.~(\ref{Igen1}) for the MI, one arrives at
Eq.~(\ref{MIt}) for the averaged MI $\tilde{\Theta}$. 
This approach is physically close to  
that used in Ref.~\cite{AK78} where
 the off-diagonal correlation term,
$\propto \d t\d \varphi$, has been neglected but 
for small rotational frequencies $\omega$, the 
gravitational field is approximated by that of the 
three-dimensional 
oblate spheroidal deformations.

Notice also
 that the density $\rho$ and energy density $\epsi(\rho)$
in Eq.~(\ref{MIt}) for the MI
within the leading leptodermic
approximation over the parameter $a/R$ is 
independent of $\alpha^{}_K$
at linear order over $\overline{\omega}$
($\alpha^{}_K \propto \overline{\omega}\propto
\omega$); see Eq.~(\ref{OmegaK}).
 As shown above, Eq.~(\ref{MI}) essentially depends
    on the off-diagonal $t,\varphi$ correlation term,
    $\mitf_{t\varphi}$, proportional
    to $\alpha^{}_K$.

  \subsection{Gravitational metric for deformed rotating NSs }
  \label{DRNSM}

For calculations of the correlation
contribution $\mitf_{t\varphi}$, Eq.~(\ref{MItf}), one has to calculate
the $\tau(r,\theta)$ factor which is a function of the radial and
polar-angle
variables \cite{RK63}. Taking into account the deformation of
the gravitational field through $g^{}_{03}(r,\theta)$, 
due to the NS rotation within 
the LPT approach on the basis of the same Schwarzschild metric solution
and assuming the azimuthal $\varphi$ invariance of the GRT
    solutions,
one will look for the gravitational field  $g(r,\theta)$ in the following form
\cite{RK63,BL67}:
\bea\l{gmnW}
&g^{}_{00}=e^\nu,~~~g^{}_{11}=-e^\lambda,~~~
g^{}_{22}=-r_{}^2,~~~g^{}_{33}\nonumber\\
&=-r_{}^2\sin_{}^2\theta~,~~~
g^{}_{03}=g^{}_{30}= 2 \tau(r,\theta) \alpha^{}_K \sin^2\theta, 
\eea
and  $g^{}_{\mu\nu}$ for all other $\mu$ and $\nu$
subscripts are zeros.
Here, $\nu=\nu(r)$, $\lambda=\lambda(r)$ 
    are known as the Schwarzschild
gravitational metric (see, e.g., Refs. \cite{KS16out,RT87,LLv2} at zero order
($\overline{\omega}=0$), and $\tau=\tau(r,\theta)$
is unknown function of the radial $r$ and angle $\theta$
coordinates in the spherical
coordinate system $t,r,\theta,\varphi$ [see Eqs.~(\ref{AWmetgen}), and
  (\ref{BLm}) for $r>R$
  and (\ref{inkerrmet}) for $r<R$]. In Eq.~(\ref{gmnW}),
$\alpha^{}_K\propto \overline{\omega} $ 
is the Kerr rotational parameter, Eqs.~(\ref{OmegaK}) and 
    (\ref{rotpar}).
Calculating the Cristoffel and Ricci tensors 
    (see Refs.~\cite{LLv2,MM26npa}),
one obtains the linear
partial-derivative equation over $\tau(r,\theta)$ by the second order:
\bea\l{eqtau03rt}
&\sin\theta~\tau(x,\theta)\left[2x^2\nu^{\prime\prime}+x^2 (\nu^\prime)^2
-x \lambda^\prime\left(4+ x \nu^\prime\right)+4\right]\nonumber\\ 
&-6 e^\lambda~\cos \theta~\frac{\partial \tau(x,\theta)}{\partial \theta}
-\sin \theta \left[
  2 e^\lambda~\frac{\partial^2 \tau(x,\theta)}{\partial
    \theta^2}\right.\nonumber\\
&- \left.x^2  \left(\lambda^\prime+\nu^\prime\right)
\frac{\partial \tau(x,\theta)}{\partial x}-
    2\frac{\partial^2 \tau(x,\theta)}{\partial x^2}\right]=0~,
\eea
where $x=r/R_{\rm S}$ for $r\leq R$, or $x=r/r_g$ for $r\ge R$)
[see Eqs.~(\ref{RS}) for the Schwarzschild radius $R_{\rm S}$ and
(\ref{rg}) for the gravitational radius $r_g$], and
$\tau(r,\theta)$ is identically re-written as $\tau(x,\theta)$ in terms of the
dimensionless variable $x$, for simplicity. 
The variables $r$ and $\theta$
in Eq.~(\ref{eqtau03rt}) are separated in the following form:
\be\l{separvar}
\tau(x,\theta)=\mathcal{X}(x)\mathcal{Y}(\theta).
\ee
%

1.~{\bf The angle solution.}
For $\mathcal{Y}(\theta)$ one finds an ordinary linear equation
of the second kind, common for outer ($r>R$) and inner ($r<R$) regions:
\be\l{Yeq}
\mathcal{Y}^{\prime\prime}(\theta)+3\cot \theta~\mathcal{Y}^{\prime}(\theta)
- q~\mathcal{Y}(\theta)=0~.
\ee
This equation for $\mathcal{Y}(\theta)$ can be solved by using
the standard textbooks.

Changing the variables by $\psi=\theta+\pi/2$,  $u=\sin(\psi)$ and
$w=(1+u)/2$, one obtains the  Hypergeometric
(Gauss) equation. 
 The non-singular solution of this equation is given by  
\be\l{solFfin}
\mathcal{Y}(w)=C_1 F\left(\alpha,\beta,\gamma,w\right),~~~~w=\cos^2(\theta/2),
\ee
where
\be\l{coefsHE}
\alpha=\frac{3+\sqrt{9-4q}}{2},~~~
\beta=\frac{3-\sqrt{9-4q}}{2},~~~
\gamma=2~.
\ee
For $q>9/4$ the hypergeometric function can be related to the conic
Mehler-Dirichlet Legendre function, $P^{\mu}_{-1/2+ip^{}_0}(\mbox{z})$,
\bea\l{solFfinMDL}
&\mathcal{Y}(w)=\frac{C_1}{\sqrt{2(1-w^2)(1+w)}} P^{-1}_{-1/2+ip^{}_0}(1-2 w),
\nonumber\\
&p^{}_0=\sqrt{4q-9}~, ~~~ q > 9/4~.
\eea
Notice also that $g_{03}(r,\theta)$, Eq.~(\ref{gmnW}),
is convergent near the edges
of the polar angles interval $[0,\pi]$. Our GRT solution for the polar angle
function $\mathcal{Y}(\theta)$ analytically differs from the Regge-Wheeler
assumption $P_\ell(\cos\theta)$ for polynomial Legendre function
($\ell$ is integer in Ref.~\cite{TR57}) for the axial symmetry case. 

2.~{\bf Outer radial solutions ($r>R$).}
The equation for the radial solution $\mathcal{X}(x)$ in the
outer case, $r>R$ and $x=r/r_g$, according to Eq.~(\ref{nulamout}), 
can be simplified because
 the coefficient in front of
 the derivative, $\partial \tau/\partial x$ term, Eq.~(\ref{eqtau03rt}), 
 is zero, $\nu^\prime+\lambda^\prime=0$. Therefore,
 one has 
\be\l{Xeq}
a_2(x)~\mathcal{X}^{\prime\prime}(x) + a_0(x,q)~\mathcal{X}(x)=0~,
\ee
where
\bea\l{aXeq}
&a_2(x)=x^2\exp\left[-\lambda(x)\right]~,\quad
a_0(x)=2 x^2\lambda^{\prime\prime}(x)\nonumber\\
&-x \lambda^{\prime}(x)\left[
  x-4-x \lambda^{\prime}(x) \right]+4+q,
\eea
and $q$ is the same variables separability constant as in Eq.~(\ref{Yeq}).
Indeed, for $\mathcal{X}(x)$ at $r>R$, one obtains the linear ordinary
second-order equation:
\be\l{Xeqout}
x^2(x-1)\mathcal{X}^{\prime\prime}(x)+\left(2-2x+xq\right)\mathcal{X}(x)=0~.
\ee
 For the limit $q\rightarrow 0$
one has
\be\l{Xeqoutq0}
 x^2\mathcal{X}^{\prime\prime}(x)-2\mathcal{X}(x)=0~.
\ee
In this limit, one obtains the same finite spherical solutions considered
in Refs.~\cite{AM25npae,MM26npa}.

\begin{figure}[tp]
          \centering
            \includegraphics[width=8.0cm]{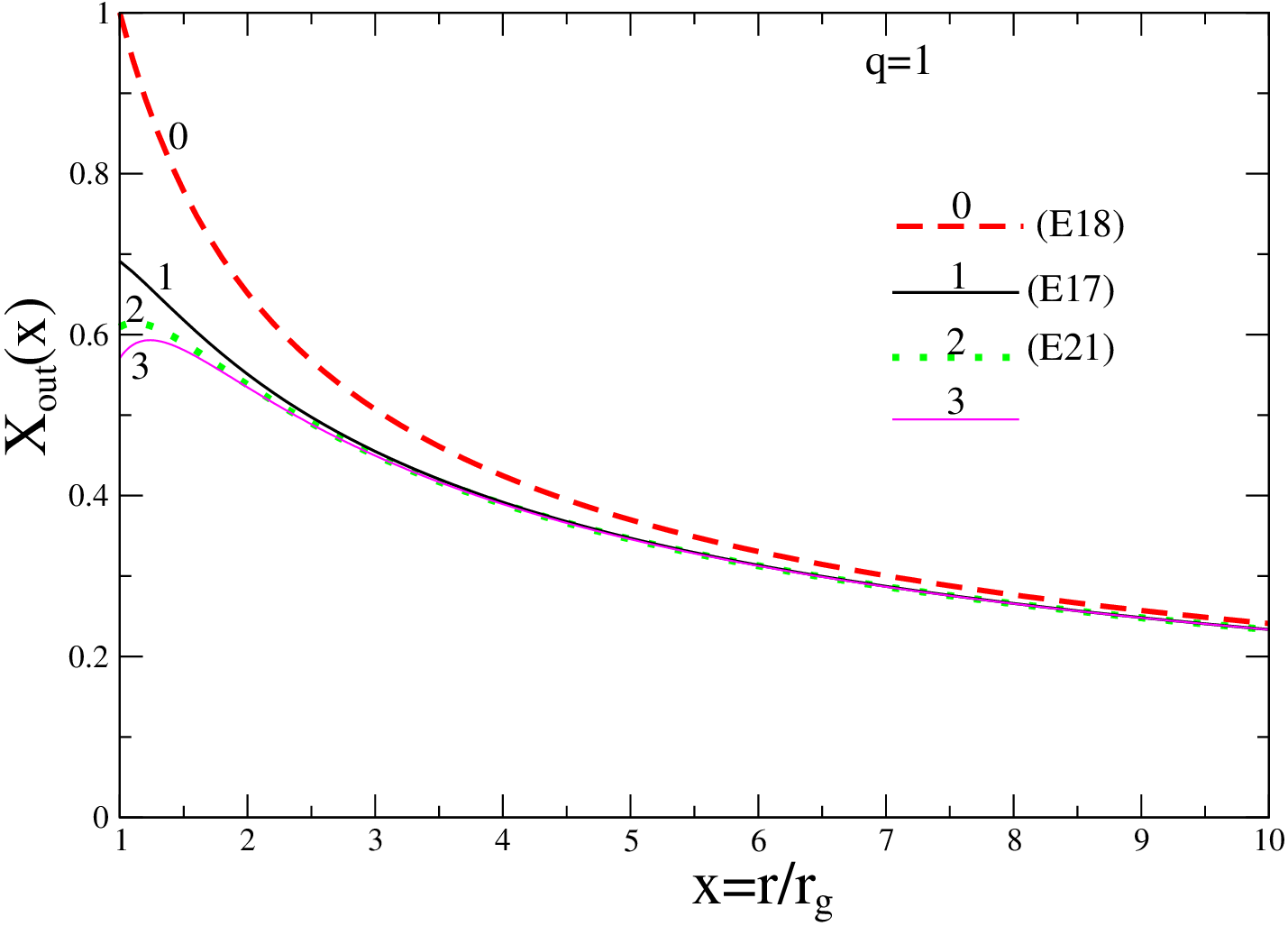}
     
  \caption{{\small 
      Asymptotic solutions $X_{out}(x)$ to the equation (\ref{Xeqout})
      for the external range $x=r/r_g>1$: ``0'' (red dashed) is the leading
      approximation with the zero order polynomial (over $1/x$ expansion,
      constant) multiplier
      to the power coefficient of the standard 
          power series method; see Appendix \ref{appC});
      ``1'' (black solid ), ``2'' (green dots) and ``3''
      (magenta solid) are related to the $n$th polynomial, $n=1, 2$, and $3$. 
      }} 
\label{fig1}
     \end{figure}

For arbitrary $q$ constants, the solutions of
Eq.~(\ref{Xeqout}) look rather complicate if we take into account general
inner solutions and the boundary condition, which connects outer and inner
solutions \cite{RT87},
\be\l{boundcond1}
\frac{r_g}{R} \approx \frac{R^2}{R^2_{\rm S}}~.
\ee
However, according to this boundary condition, for the case of
$R\ll R_{\rm S}$, for which one finds simple analytical inner solutions for
the $\mathcal{X}(x)$, we need
only asymptotic outer solutions for large $x=r/r_g$ (see Fig.~\ref{fig1} and
its discussions around it).

These asymptotic outer solutions for $x=r/r_g \gg 1$ can be obtained by using
the expansion up to definite order $n$ terms over $1/x$
analytically
by the standard power series method (PSM);
see Appendix \ref{appC}.
 Up to the first order $n=1$,
 one obtains the following convergent result $\mathcal{X}_1(x,q)$
 for the outer solution:
 \bea\l{Xout2}
&\mathcal{X}_1(x,q)= C_1x^{(1-\sqrt{9-4q})/2} 
 \left[1+\frac{q\left(1-\sqrt{9-4q}\right)}{4(2-q)x}\right.\nonumber\\
   &+\left. O\left(\frac{1}{x^2}\right)\right],~~~q~<~9/4~,
\eea
where the constraint on $q$ is just
opposite with that in Eq.~(\ref{solFfinMDL}). Therefore, we should
exclude the conic Mehler-Dirichlet Legendre functions from our results.
For the leading term, from Eq.~(\ref{Xout2}), one finds
\be\l{Xout1}
\mathcal{X}_1(x,q)= C_1x^{(1-\sqrt{9-4q})/2}\left[1+O\left(\frac{1}{x}\right)
  \right].
\ee

 Similarly, one finds
the outer solutions for any polynomial orders $n$.
    Notice that for $q=0$, our solution coincides
    with the Boyer\&Linquist result \cite{BL67,AM25npae,MM26npa}
    by using $C_1=1$ for
        arbitrary constant $C_1$.

   Figure \ref{fig1} shows fast convergence of the external ($x>1$)
 solution to the equation (\ref{Xeqout}) with increasing the expansion
 power $n$. As shown above, even the leading approximation for $n=0$ can be used
 for the solution derivations 
 with good accuracy.
  For large values of $x$,
 $x=r/r_g\simg 3.5$, even for $n=1$, one obtains
 good convergence. For $n=3$, 
 one finds a good convergence for $x\simg 1.5$, and so on. Thus, practically,
 one has the outer solution for any finite distance from $x=1$, taking the
 larger
 $n$ the smaller this distance by using Mathematica. However, as seen from
 Figure \ref{fig1} and the analytical inner $(r<R)$ results
 for $r \ll R_{\rm S}$
 (see the next section), one can stop
 even at $n=0$ 
 with accuracy $7$ \% for
 large values of $r/r_g\simg 5$.
 
  3.~{\bf Inner solutions ($r < R$).}
 Substituting Eq.~(\ref{nulamin}) into Eq.~(\ref{aXeq}), one obtains
   from Eq.~(\ref{Xeq}) the equation for $\mathcal{X}(x)$
   in the inner region, $r<R$,
   \be\l{Xeqinx}
   a^{}_2(x)\mathcal{X}^{\prime\prime}(x)+a^{}_1(x)\mathcal{X}
   ^{\prime}(x)+a^{}_0(x,q) \mathcal{X}(x)=0~,
   \ee
   where
   \begin{widetext}
   \bea\l{acoefrt}
   &a^{}_0 =
   2\left[6 \sqrt{1-\xi^2}-\left(10-9\xi^2\right)\sqrt{1-x^2} 
    +x^4\left(21\sqrt{1-\xi^2}-4\sqrt{1-x^2}\right)\right.
     \nonumber\\
     & -\left. x^2\left(27\sqrt{1-\xi^2} 
     -\left(32-27\xi^2\right)\sqrt{1-x^2}\right)
     +q\left(-6 \sqrt{1-\xi^2}+\left(10-9\xi^2\right)\sqrt{1-x^2}
   +x^2\left(6 \sqrt{1-\xi^2}-\sqrt{1-x^2}\right)\right)\right],
   \nonumber\\
   &a^{}_1 =3x^3\left[\sqrt{1-\xi^2}\left(1-x^2\right)
      -3\left(1-\xi^2\right)\sqrt{1-x^2}\right],~~~
   a^{}_2 =x^2(x^2-1)\left[6\sqrt{1-\xi^2}-
     \left(10-9\xi^2\right)
     \sqrt{1-x^2}\right.\nonumber\\
     &-\left.x^2\left(6\sqrt{1-\xi^2}
    -\sqrt{1-x^2}\right)\right],\quad  \xi=R/R_{\rm S}~.
   \eea
   \end{widetext}
              To simplify Eq.~(\ref{Xeqinx}) for analytical solutions,
          we expand
   the coefficients, Eq.~(\ref{acoefrt}), over power series, first over small
   $\xi$, and then, therefore ($r\leq R$, or $x\leq \xi$) over small
   $x=r/R_{\rm S}$ , both up to quadratic terms. Finally, one obtains
   from Eq.~(\ref{acoefrt}) 
   \bea\l{acoefexp}
   &a^{}_0(x,q) =(q-2)\left(4-6\xi^2\right)+20x^2,\nonumber\\
     &  a^{}_1(x) =0~,\qquad a^{}_2(x) =4x^2~.
   \eea
   With these coefficients,
   one arrives at
   much simpler equation:
   \be\l{14tauinexp}
   x^2\mathcal{X}^{\prime\prime}(x)
   +(q-2)\left(1-\frac{3}{2} \xi^2\right) \mathcal{X}(x)=0~.
\ee
This is the Bessel equation which has
the following non-singular solution: 
\be\l{14tauinexpsol}
\mathcal{X}(x)=c^{}_1\sqrt{x}J_{p}\left(\sqrt{5}x\right),
\ee
but with the modified index $p=p(\xi,q)$,
\be\l{prt}
p=\frac12\sqrt{9-4q-6\xi^2\left(2-q\right)}~.
\ee
In Eq.~(\ref{14tauinexpsol}),
$c{}_1$ is still an arbitrary integration constant.

The coefficient
$\tau$ 
of $g^{}_{03}$
       as function of  the
       radial,  $x=r/R_{\rm S}$, and angle, $y=\theta$,
       $\sin^2\theta=\varrho^2/r^2$, 
       coordinates (Appendix \ref{appD}),
       is smoothly depending on $\xi$  up to $\xi \approx 0.5$, for a small
       separation constant $q$.       
          The coefficient $c^{}_1$ in Eq.~(\ref{14tauinexpsol})
       is determined  by the
       surface boundary condition, Eq.~(\ref{boundcond1}),
       at the ES $r=R$ for $q<9/4$,
          \be\l{resc1}
       c^{}_1=\frac{\mathcal{X}_1\left(\frac{1}{\xi^2}\right)}{\sqrt{\xi}
         J_p\left(\sqrt{5}\xi\right)},\qquad \xi=\frac{R}{R_{\rm S}}\approx
         \sqrt{\frac{r_g}{R}}~.
       \ee
       Generally speaking, the separation constant $q$ 
           places the role of an additional
       integral of motion beyond the energy and angular momentum under a
       gravitational field.

       Thus, instead of the vacuum solutions of the GRT equations
       \cite{TR57} we found
       the outer ($r>R$) and the inner ($r<R$) solutions of the corresponding
       equations with the boundary condition for their continuous transition
       through the finite surface radius $r=R$.
     
\section{ADIABATIC MI APPROACH}
\l{SFA}

\subsection{Macroscopic NS moments of inertia}
\l{NSMI}

For small frequencies of the 
macroscopic NS rotation,
one should expect mostly an adiabatic 
     motion. In our ESA, it obeys
the condition of smallness
of the rotation energy, $E_{\rm rot}$, with 
respect to the ETF energy,
$E$, Eq.~(\ref{enerden}),
\be\l{adcond}
E_{\rm rot} \ll E~,~~~~~E_{\rm rot} = \frac{1}{2}\Theta \omega^2~.
\ee
%
%
%
This condition was studied in details in Refs.~\cite{MM26npa,AM25npae} 
    for Hogan's and our spherical
off-diagonal gravitational metric coefficient $\tau(r)$. 
Under the adiabatic condition, Eq.~(\ref{adcond}),
the MI $\Theta$, Eq.~(\ref{MI}), can be evaluated in terms of the
energy density $\epsi(\rho)$, Eq.~(\ref{enerden}), by using the
statistically averaged component
$\tilde{\Theta}$, Eq.~(\ref{MIt}), and
its  off-diagonal $t-\varphi$ correction 
$\mitf_{t\varphi}$, Eq.~(\ref{MItf}).
For this purpose, in the linear
approximation over $\overline{\omega}$, Eq.~(\ref{rotpar}),
for the angular momentum
$I$,  i.e., zero approximation for the MI $\Theta$,
we neglected
all quadratic
terms of the order of $\overline{\omega}^2\propto\alpha_K^2$,
in the
volume element, $\d \mathcal{V}_\omega\rightarrow
\d \mathcal{V}$, Eq.~(\ref{dV}). 
Splitting now $\epsi(\rho)$, Eq.~(\ref{enerden}),
into non-gradient, $\mathcal{A}(\rho)$,
Eq.~(\ref{vol}),
and gradient,
$\mathcal{C}(\nabla \rho)^2\approx \mathcal{C}(\partial \rho/\partial r)^2$,
dependent components,
and using similar techniques as those 
in Refs.~\cite{MM24,MM26npa},
one can present each of  them,
$\tilde{\Theta}$ and $\mitf_{t\varphi}$, in terms of
the volume and surface
terms,
\be\l{adMIVS}
\tilde{\Theta}=\tilde{\Theta}_V + \tilde{\Theta}_S~, \qquad
\mitf_{t\varphi}=\mitf_V + \mitf_S~.
\ee
Here, $\tilde{\Theta}_V$ and $\mitf_V$ are the volume, and
$\tilde{\Theta}_S$ and  $\mitf_S$  are
the surface components of the
statistically averaged MI and its $t,\varphi$
correction, respectively (see Ref.~\cite{MM26npa}).

For a small  angular momentum
${\textit {\bf I}}$, one can direct the
axially symmetry axis $z$ to the vector ${\bf I}$, and introduce the
spheroidal coordinate system 
with this $z$ axis.
Due to such a rotation, one has a
deformation
of the surface NS shape in the effective surface leptodermic
approximation. 
In this case, one can use for the finite perfect dense-fluid expression
(\ref{Igen}) for  ${\bf I}$ with the energy-momentum tensor $T_{\mu\nu}$
(\ref{Tmn}) and  four-velocity
$u_\mu=u^t(1,0,0,\overline{\omega})$. Thus, one obtains with the help of
Appendix \ref{appB} the same MI expresions $\Theta$, Eq.~(\ref{MI}),
in terms of the
statistically averaging $\tilde{\Theta}$, Eq.~(\ref{MIt}),
and correlation $\mitf_{t\varphi}$, Eq.~(\ref{MItf}),
MI components. However,  $\tau(r,\theta)$ in Eq.~(\ref{MItf}) is now
the solution of the
equation (\ref{eqtau03rt}) for $\tau $ as function of the radial  and
angle variables in a separable form (\ref{separvar}), where
$\mathcal{X}(r/R_{\rm S})$
inside of the NS are given by Eqs.~(\ref{14tauinexpsol}) and (\ref{prt})
with Eq.~(\ref{resc1}) for the constant $c^{}_1$, 
and
$\mathcal{Y}(\theta)$ is presented by Eq.~(\ref{solFfin}). 
    Therefore, one again
obtains the additional constraint, $r_g<R<R_{\rm K}<R_{\rm S}$, due to the
consistent NS  rotation under a strong gravity,
where $R_{\rm K}$ is the pole of the MI, Eq.~(\ref{MI}).
%
%
\begin{figure}[htp]
  \begin{center}
    \includegraphics[width=0.5\textwidth,clip]{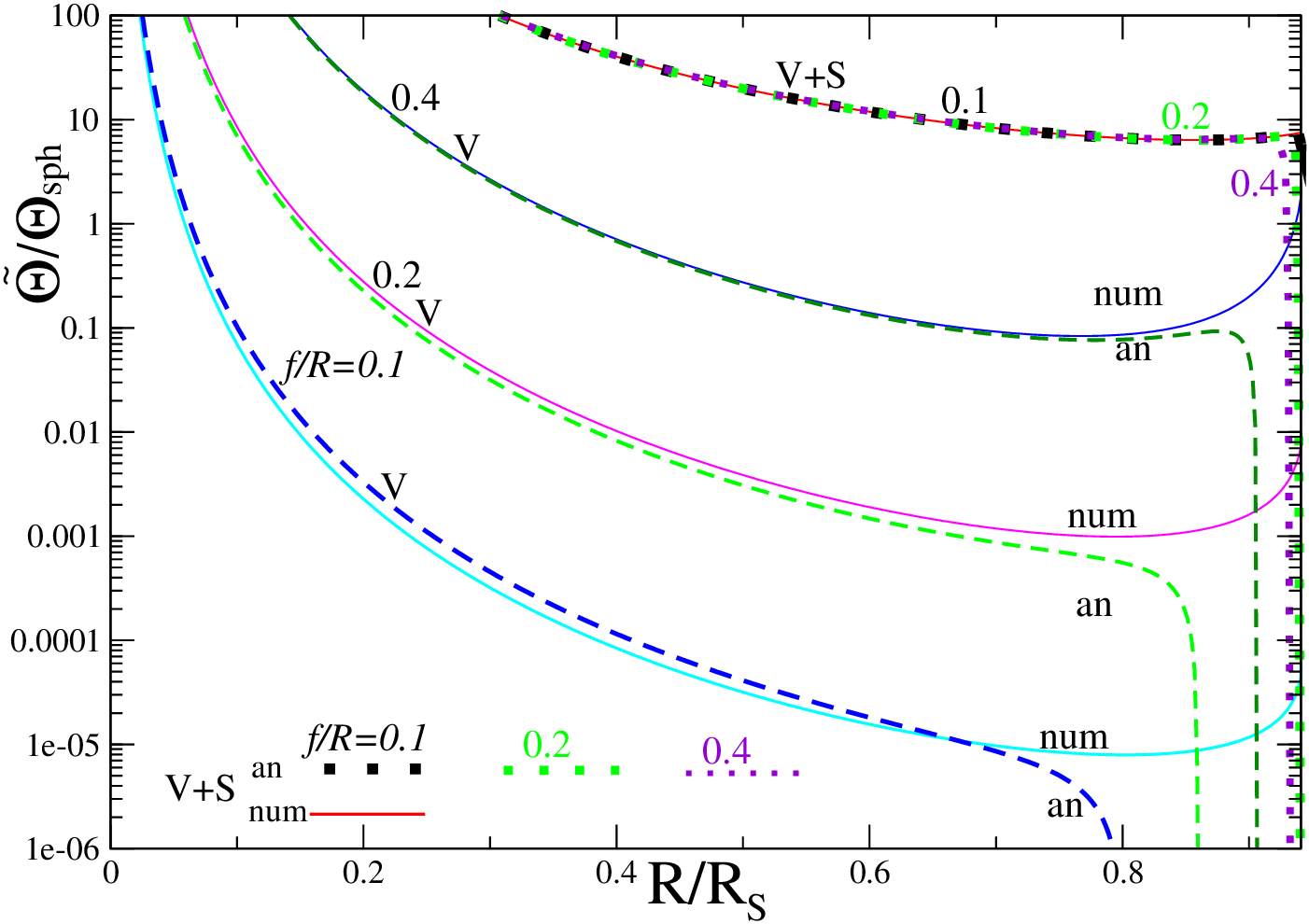}
    \end{center}
  
  \caption{{\small The full (volume and surface, ``V+S'')
      statistically averaged contributions
      $\tilde{\Theta}$, Eq.~(\ref{adMIVS}), in units of
      the spherical MI solid value $\Theta_{\rm sph}=2MR^2/5$, where
      $M$ is the NS mass and $R$ is the effective NS radius, are shown.
      They are presented with the numerical,
      Eq.~(\ref{MIeq2appF}), versus analytical,
      Eq.~(\ref{MIeq4appF}), approximations  for the
      integral $\tilde{I}$ in the volume component
      $\tilde{\Theta}_V$, Eq.~(\ref{MIeq1appF}),
      as functions of the radius variable,
      $R/R_{\rm, S}$. The different foci distances are
      small: 
      $\overline{f}={\it f}/R=0.1$, $0.2$, and
      $0.4$, 
      which are related to the corresponding
      spheroid deformations
      $\mathcal{Q}=\mbox{b}/\mbox{a} \approx 1-{\it f}$ up to second
      order terms
      [see Eq.~(\ref{MIeq3appF})].        
       }}
\label{fig2}
\end{figure}
The basic reason of appearance of this constraint is the
    off-diagonal
gravitational metric $g_{03}$ 
parameter which is approximately  proportional,
in a self-consistent way, 
to the angular momentum $I$ by
Eq.~(\ref{OmegaK}),
$\alpha^{}_K \propto \overline{\omega} \propto I$,
for a small rotational frequency 
parameter $\overline{\omega}$;
see Appendix \ref{appB}.

\subsection{Volume statistically averaging component}
\l{appF1}

For the volume statistically averaged component $\tilde{\Theta}$,
Eq.~(\ref{MIt}), one has
\be\l{MIeq1appF}
\tilde{\Theta}
=\frac{15\Theta_{\rm sph}}{2}\left(\frac{f}{R}\right)^5
\tilde{\mathcal{I}},
\ee
where $f=\sqrt{\mbox{a}^2-\mbox{b}^2}$ is the distance between the
two spheroid foci,
$\mbox{a}$ and $\mbox{b}$ are the oblate spheroid semiaxises,
$\mbox{a}>\mbox{b}$~.
In Eq.~(\ref{MIeq1appF}),
\bea\l{MIeq2appF}
&\tilde{\mathcal{I}}=
\int_{0}^{\zeta_b} {\rm d}\zeta \int_{0}^{1}{\rm d}\eta\nonumber\\
&\times
\frac{\left(\zeta^2+\eta^2\right)\left(1+\zeta^2\right)\left(1-\eta^2\right)}{
  \sqrt{1-\overline{f}^2\xi^2\left(1+\zeta^2-\eta^2\right)}
  \left[A_{\rm S}-\sqrt{1-\overline{f}^2\xi^2\left(
        1+\zeta^2-\eta^2\right)}/2\right]^2}~,
\eea
where $\overline{f}=f/R$, and $A_{\rm S}$ is given by Eq.~(\ref{AS}).
The upper integration limit $\zeta_b$ in the integration over $\zeta$
is determined by the deformation parameter
$\mathcal{Q}=\mbox{b}/\mbox{a}<1$, as
\be\l{MIeq3appF}
\zeta_b={\rm arctanh}\mathcal{Q}=
     {\rm arctanh}\left(\frac{\mbox{b}}{\mbox{a}}\right).
\ee
Using the volume conservation condition $\mbox{a}^2\mbox{b}=R^3$
for the dense NS matter,
one has the only one deformation parameter $\mathcal{Q}$,
$\mbox{a}=R\mathcal{Q}^{-1/3}$ and
$\mbox{b}=R\mathcal{Q}^{2/3}$ [it is not in fact
necessary to use this condition but having instead of
one deformation parameter $\mathcal{Q}$ the two
constants $\mbox{a}$ and $\mbox{b}$, $\overline{f}\approx \sqrt{2(1-Q)}$].
Expanding the integrand of the expression (\ref{MIeq2appF}) over the focus
distance in units of the NS radius, $f/R$, up
to second order, one can
explicitly take the double integral,
\bea\l{MIeq4appF} 
&\tilde{I}=\frac{8 \zeta_b\left(1 + \zeta^2_b\right)^2}{15(2 A_{\rm S}-1)^3}
\left[2A_{\rm S} + \frac{2}{7}\overline{f}^2\xi^2 \left(1 + \frac{5}{4} \zeta^2_b
  \right)
  - 1\right.\nonumber\\
  &-\left. \frac{6}{7} \overline{f}^2\xi^2
\left(1 +\frac{5}{4} \zeta^2_b\right)\right].
\eea
\subsection{Volume correlation component}
\l{appF2}

Substituting the inner solution (\ref{separvar})  for $\tau$  with
(\ref{14tauinexpsol}) for $\mathcal{X}(r/R_{\rm S})$
and (\ref{solFfin}) for $\mathcal{Y}(\theta)$ into Eq.~(\ref{MItf})
for the correlation MI component, one finds
\bea\l{MIeq5appF}
&\mathcal{T}_{t\varphi}=
\frac{6 c^{}_1\overline{f}^{7/2}}{\xi^3}~
\mathcal{I}_{t\varphi}~,\qquad \overline{f}=\frac{f}{R}~,
\eea
\begin{widetext}
\bea\l{MIeq6appF}
&\mathcal{I}^{}_{t\varphi}=
\int_0^{\zeta_b} \int_{0}^{1}\frac{{\rm d}\zeta{\rm d}\eta}{
  \left(1+\zeta^2-\eta^2\right)^{3/2}}
\frac{\left(\zeta^2+\eta^2\right)\left(1+\zeta^2\right)\left(1-\eta^2\right)}{
  \sqrt{1-\overline{f}^2\left(1+\zeta^2-\eta^2\right)}
  \left[A_{\rm S}-\sqrt{1-\overline{f}^2\left(1+\zeta^2-\eta^2\right)}/2\right]^2}
\nonumber\\
&\times
J_p\left(\sqrt{5}~\overline{f}\sqrt{1+\zeta^2-\eta^2}\right)
\, _2F_1\!\left[\!\frac{3\!+\!\sqrt{9\!-\!4q}}{2},\!\frac{3\!-\!\sqrt{
      9\!-\!4q}}{2};\!2;\!\frac{1}{2}\left(\!1\!+\!\frac{\zeta\eta}{\sqrt{
      \!1\!+\!\zeta^2\!-\!\eta^2}}\right)\right].
\eea
\end{widetext}
In order to obtain the analytical approximate results, 
one can first try to fix the constant index $p(\xi,q)$
of the Bessel function $J_p$, Eq.~(\ref{prt}),
and $\alpha(q)$, $\beta(q)$ arguments
of the Hypergeometric
function $\, _2F_1$. Then, we may do the power expansions
of the Bessel, Eq.~(\ref{14tauinexpsol}),  and hypergeometric,
Eq.~(\ref{solFfin}), functions over their arguments for a
fixed $p$, $\alpha$,
$\beta$ ($\gamma=2$), Eq.~(\ref{coefsHE}), and $A_{\rm S}$, Eq.~(\ref{AS})
 for $\xi \ll 1$.
Expanding then the result over
the deformation
parameter $\overline{f}$, one can simplify the expression
for $\mathcal{I}^{}_{t,\varphi}$, Eq.~(\ref{MIeq6appF}). This double integral
can be integrated in terms of the Gauss hypergeometric functions:
$\mathcal{I}^{}_{t,\varphi}=\mathcal{F}(\zeta_b)$, where
\begin{widetext}
\bea\l{MIeq7appF}
\mathcal{F}(z)&=&
\frac{2^{-p} 5^{p/2-1} z^{1 - p}\left(z\overline{f} R/R_{\rm S}\right)^p}{21p (1 - 2 A)^2
  (p^2-4) \Gamma(1 + p)}
\left\{-7 p (12 - 4 p - 3 p^2 + p^3) (4 + \alpha \beta)
  \right.\nonumber\\
&\times&\left.
   \, _2F_1\left(\frac{1}{2}, \frac{1 - p}{2}, \frac{3}{2}, -z^2\right)
    +2 p (p^2-4) (4 + \alpha \beta) \left[2 (p-3) + 7\right]
    \, _2F_1\left(
    \frac{1}{2}, \frac{3-p}{2}, \frac{3}{2}, -z^2\right)
\right.\nonumber\\
    &+&\left. 
7 z \left[15 \alpha \beta
(1 + z^2) \left(2 z^p - 2 (1 + z^2)^{p/2} + 
         p (1 + z^2)^{p/2} + 
         2 z^2 \left(z^p -(1 + z^2)^{p/2}\right)\right)
\right.\right.\nonumber\\
    &+&\left.\left. 
         2 p z (p^2-4) (4 + \alpha \beta) \left[
           2~ \, _2F_1\left(\frac{3}{2}, \frac{3 - p}{2},
           \frac{5}{2}, -z^2\right) + 
         z^{2} 
     \, _2F_1\left(\frac{5}{2}, \frac{3 - p}{2}, \frac{7}{2}, -z^2\right)
     \right]\right]\right\},
\eea
\end{widetext}
         where $p$ is the Bessel function index, Eq.~(\ref{prt}), and
         $\alpha$ and $\beta$ are the coefficients of the hypergeometric
         functions $\, _2F_1(\alpha,\beta; 2; w)$; see Eq.~(\ref{coefsHE}).

\subsection{Surface statistical and correlational components}
\l{appF3}

For the surface parts,
$\tilde{\Theta}_S$ and $\mathcal{T}_{t\varphi}^{(S)}$,
one can reduce them
to the simple one-dimensional integrals. They can be analytically
expressed in terms of the tension coefficient $\sigma \propto a/R$
as the surface energy;
see Appendix \ref{appA}.
For the statistical surface part
$\tilde{\Theta}_S$, with the simplest interparticle interaction
$\mathcal{C}$ for small oblate  defomations, one finds
\be\l{MItS}
\tilde{\Theta}_S=\frac{\mathcal{C}}{c^2}\int {\rm d} \mathcal{V} J(r)
\varrho^2 e^{-\nu_{\rm in}} (\nabla \rho)^2~,
\ee
where $\varrho$ is the cylindrical variable,
$\varrho,\mbox{z},\varphi$, related to the oblate spheroidal coordinates
by the standard equations. 
Similarly, for the correlation component
 $\mathcal{T}_{t\varphi}^{(S)}$ , one arrives at
\be\l{MItfS}
\mathcal{T}^{(S)}_{t\varphi}=
\frac{\mathcal{C}}{Mc^2}\int {\rm d} \mathcal{V} J(r)
\varrho^2 e^{-\nu_{\rm in}} (\nabla \rho)^2~\frac{\tau}{\varrho^2+\mbox{z}^2}~.
\ee
Here, $J$ is the same inner gravitational volume-shortness
Jacobian, Eq.~(\ref{Jin}), $\nu_{\rm in}$ is the 
inner Schwarzschild metric function, Eq.~(\ref{nulamin}), and
$\tau$ was derived above; see
Eq.~(\ref{14tauinexpsol}) for $\mathcal{X}(r/R_{\rm S})$
and (\ref{solFfin}) for $\mathcal{Y}(\theta)$ for the inner separable
solution, Eq.~(\ref{separvar}). 
    In Eq.~(\ref{MItfS}), the integral over 
${\rm d} \mathcal{V}$ is definite as taken over the spatial variables
    and  the inner Jacobian $J$ accounts for a strong gravitation for
    the Schwarzschild metric.

Thus, from Eqs.~(\ref{MItS}) and (\ref{MItfS}), following
Appendix \ref{appD},
one arrives at
\be\l{adMItS}
\frac{\tilde{\Theta}_S}{2 MR^2/5}=\frac{10\pi}{3}~\frac{\sigma R^2}{Mc^2}
~\frac{1}{1-R^2/R_{\rm S}^2}~,
\ee
and
\be\l{adMItfS}
\mathcal{T}_{t\varphi}^{(S)}
=\frac{4 \pi}{3}\frac{\sigma R^2}{Mc^2} \frac{r_g}{R} =
\frac{8\pi \sigma R^2 G}{Rc^4}~.
\ee
   We neglected higher order terms $\sim (a/R)(f/R)$, 
and used the general ESA method (Appendix \ref{appD}), 
with the help of Eqs.~(\ref{eq0}) for the derivative of the density,
$\nabla \rho \approx \partial \rho/\partial r$, (\ref{enertots})
for the tension coefficient $\sigma$ [$\sigma \propto (a/R)J(R)$],
and (\ref{rg})
for the gravitational radius ($r_g\propto M$)
within the same accuracy.

All surface MI components are proportional
    to the leptodermic parameter $a/R$ through the surface tension
    coefficient $\sigma$.
As shown in 
Ref.~\cite{MM24} and
Appendix \ref{appA},
this coefficient
 can be calculated
 analytically; see Eq.~(\ref{enertots}).

 As shown in Fig.~\ref{fig2}, our numerical code,
 Eqs.~(\ref{MIeq2appF}) with (\ref{MIeq2appF}),
 and its analytical approximation (\ref{MIeq4appF}) for the integral
 $\tilde{I}$
in the calculations of
the volume statistically averaged MI component $\tilde{\Theta}$
(\ref{MIeq1appF})
converge well to each other with a good accuracy, except for small
ranges near the asymptotes. This is related to
the result of the integration
over two spheroidal coordinates $\eta$ and $\zeta$; see Eq.~(\ref{MIeq2appF}).
The ``num'' and ``an'' curves 
``$V+S$'' for the total volume and surface statistically
    averaged MI
with different deformations $\it{f}/R$ are almost identical,
except for a small doubtful areas near the asymptotes at $R/R_{\rm S}=1$;
see this figure and a warning for the ESA 
analytical derivations for the surface contributions near
the asymptotes in
Refs.~\cite{MM24,AM25npae,MM26npa}.
Notice also that one obtains a
nonmonotonic behaviour of the volume MI $\tilde{\Theta}_V$ (``V'' curves)
as function of
the variable $R/R_{\rm S}$
at all small deformations $\it{f}/R$ 
due to a strong gravitation. Moreover, we found minima, $R/R_{\rm S} \approx
0.7-0.8$, even in the statistically averaged MI
at all these deformations under the consideration.
The measure of the surface contribution, $\tilde{\Theta}_S$,
Eq.~(\ref{adMItS}), into
the full statistically averaged MI $\tilde{\Theta}$, Eq.~(\ref{adMIVS}),
is obviously a large difference between the ``$V+S$'' and ``$V$'' curves.

    Notice that in our macroscopic derivations we use the linear
           perturbation approach
           over a small dimensionless parameter
           $\overline{\omega}$, Eq.~(\ref{rotpar}),
           $\overline{\omega}\ll 1$. To clarify this approach, it
           is convenient to present also
           the so called adiabatic approximation
           in terms of a small ratio of the
           NS rotational energy to the corresponding
           non-rotating ETF energy used in the macroscopic physics. 
           Using now Eqs.~(\ref{MI}), (\ref{adMIVS})-
            (\ref{adMItfS}),
one can specify
      the adiabatic condition (\ref{adcond}) for the 
       NS rotation
            periods $P=2\pi/\omega$, 
      \be\l{omcond}
           P\gg P_0~,
      \ee
      where $P_0$ is the characteristic upper period limit
      for adiabatic motion,
      \be\l{om0}
          P_0=2\pi\sqrt{\frac{\Theta}{2E}}~.
           \ee
           Here, $\Theta$ is the full NS MI,
           Eq.~(\ref{MI}), and $E=E_{ETF}$
          is the full macroscopic energy in the non-rotating
            neutron-star (zero-order) ETF approximation, Eqs.~(\ref{energytot})
          and (\ref{enertot});
           \begin{figure}[htp]
             \begin{center}
   \includegraphics[width=0.5\textwidth,clip]{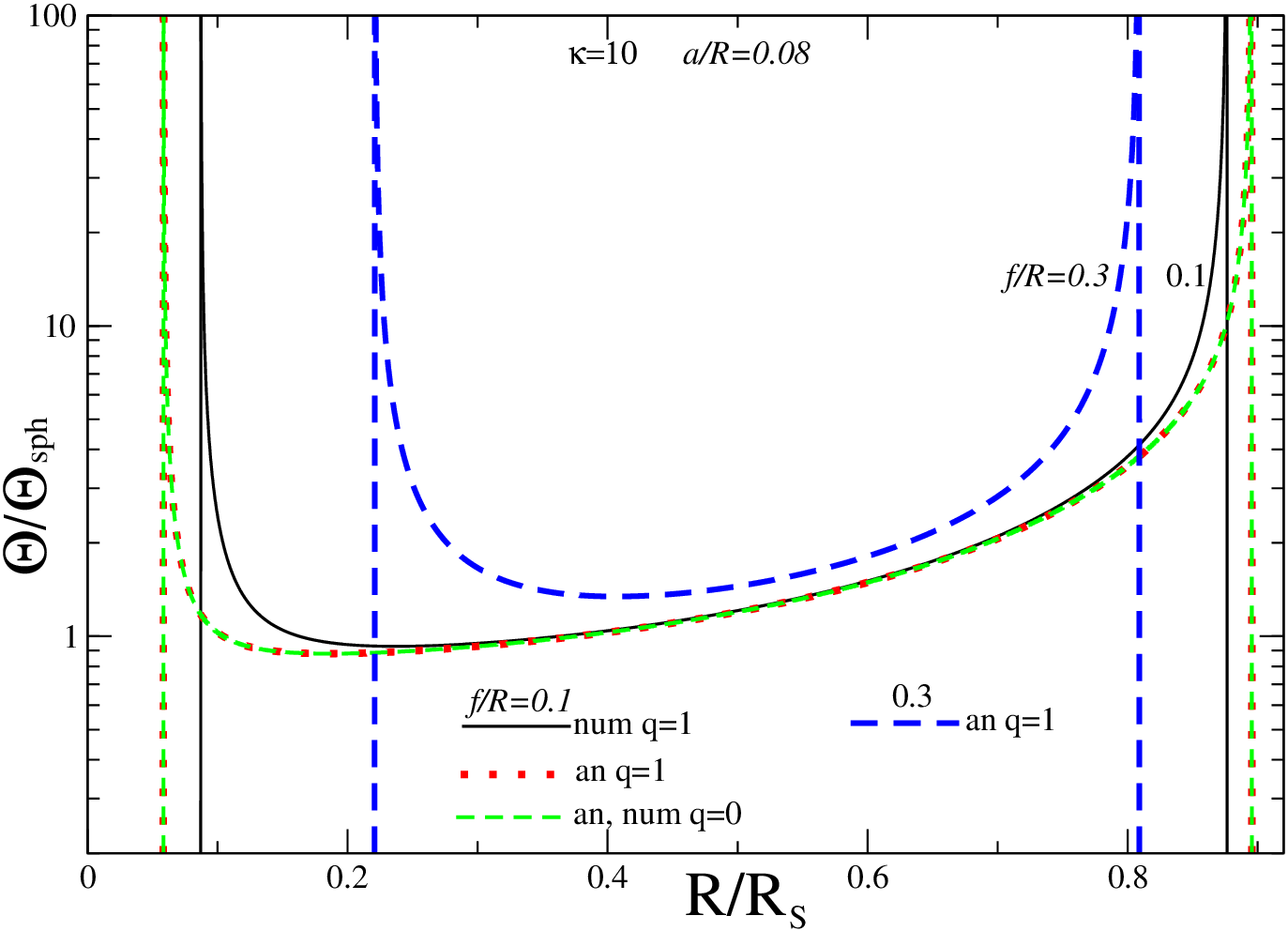}
             \end{center}

             \caption{{\small
    Moments of inertia $\Theta$, Eqs.~(\ref{MI}), (\ref{MIt}), (\ref{MItf}),
  (\ref{adMIVS})
      in units of $\Theta_{\rm sph}=2MR^2/5$
      as function of the radius in units of $R_{\rm S}$, $R/R_{\rm S}$,
      for different deformations
      $f/R=0.1$ and $0.3$;  separable constants are given by spherical
      $q=0$ and deformed $1$ cases;
      see also Eqs.~(\ref{MIeq1appF}),
      (\ref{MIeq4appF}), (\ref{MIeq1appF}), (\ref{MIeq7appF} for the volume, and
      (\ref{adMItS}) and (\ref{adMItfS}) for the surface parts. The solid line shows the  calculations
      with numerical corellation contribution, Eq.~(\ref{MIeq6appF}). 
      The relative crust thickness is $a/R=0.08$, and the dimensionless
      nuclear-gravity incompressibility $\kappa$, Eq.~(\ref{kappa}),
      is strong, $\kappa=10$.
    }}
\label{fig3}
\end{figure}
\vspace{-0.2cm}            
\begin{table*}
 \footnotesize 
  \begin{tabular}{|c|c|c|c|c|c|c|c|c|c|}
\hline
NS & $M$ & $R$ & $P$ & Refs. &$\left(\frac{Ic}{M^2G}\right)_{\rm sph}$ &$\frac{Ic}{M^2G}$
& $\frac{Ic}{M^2G}$& $\frac{Ic}{M^2G}$ &$P_0$
\\
&  ($M_{\odot}$) & (km) & (ms) &  &$10\Vert 0.03$ & $10\Vert 0.1$ & $10\Vert 0.3$
& $50\Vert 0.1 $ &$50\Vert 0.1$ (ms)
\\
\hline
J0030+0451& $1.43^{+0.20}_{-0.17}$& $12.68^{+1.31}_{-1.04}$ & $4.9$
& \cite{MM19}
&  $ {0.17;0.21\atop{0.17;0.22}}$
&  $ {0.17;0.21\atop{0.18;0.22}}$
& $ {0.21;0.25\atop{0.21;0.27}}$
& $ {2.42;1.99\atop{1.61;1.69}}$
& $ {0.21;0.19\atop{0.16;0.16}}$ 
\\
& $1.34^{+0.15}_{-0.16}$& $12.71^{+1.14}_{-1.19}$ &
&\cite{TR19}
&  $ {0.16;0.20\atop{0.17;0.23}}$
&  $ {0.16;0.20\atop{0.17;0.23}}$
& $ {0.20;0.24\atop{0.21;0.27}}$
& $ {2.14;1.84\atop{1.46;1.61}}$
&   $ {0.097;0.110\atop{0.14;0.15}}$
\\
\hline
J0740+6620& 2.072$^{+0.067}_{-0.066}$ & $12.39^{+1.30}_{-0.98}$&
2.9
& \cite{TR21}
&  $ {0.30;0.32\atop{0.28;0.32}}$
&  $ {0.30;0.33\atop{0.29;0.32}}$
& $ {0.49;0.43\atop{0.42;0.41}}$
& $ {-4.42;15.5\atop{-9.4;7.6}}$
& $ {0.23;0.46\atop{0.34;0.31}}$
\\
& $2.08^{+0.07}_{-0.07}$ & $12.76^{+1.49}_{-1.02}$ &
&\cite{AD24}
&$ {0.30;0.35\atop{0.29;0.35}}$
&$ {0.30;0.35\atop{0.29;0.35}}$
& $ {0.47;0.45\atop{0.41;0.44}}$
& $ {-5.99;7.18\atop{-19.3;5.27}}$
&$ {0.27;0.32\atop{0.48;0.26}}$  
 \\
\hline
J1731-347& 0.77$^{+0.20}_{-0.17}$ & $10.4^{+0.86}_{-0.78}$&
147
& \cite{VD22}
&${0.48;0.61\atop{0.63;0.83}}$
&${0.48;0.61\atop{0.64;0.84}}$
&${0.58;0.74\atop{0.80;1.10}}$
&${4.02;4.38\atop{3.83;4.76}}$
&${0.18;0.13\atop{0.094;0.106}}$
\\
& $0.83^{+0.17}_{-0.13}$ & $11.25^{+0.53}_{-0.37}$ & 
&\cite{VD22}
&${0.57;0.64 \atop{0.70;0.80}}$
&${0.57;0.64 \atop{0.71;0.81}}$
&${0.69;0.78\atop{0.88;1.03}}$
&${4.32;4.56\atop{4.29;4.76}}$
&${0.13;0.13\atop{0.11;0.11}}$
\\
&&&&  & \tiny{$\times 10^{-2}$} & \tiny{$\times 10^{-2}$}
&\tiny{$\times 10^{-2}$}&\tiny{$\times 10^{-2}$}
& 
\\
\hline
Cen X-3& 1.21$^{+0.21}_{-0.21}$ & $9.178^{+0.130}_{-0.130}$&
4.84
& \cite{SN11}
&${1.29;1.32\atop{1.33;1.38}}$
&${1.30;1.32\atop{1.33;1.38}}$
&${1.93;1.89\atop{1.61;1.67}}$
&${6.66;4.49\atop{1.26;1.25}}$
&${0.31;0.26\atop{0.12;0.12}}$ 
\\
& 1.49$^{+0.08}_{-0.08}$ & &
& \cite{MR11,wiki:PSR-Cen-X-3}
&${1.34;1.36\atop{1.28;1.31}}$
&${1.35;1.37\atop{1.29;1.32}}$
&${1.93;1.90\atop{1.69;1.70}}$
&${-(5.72;13.6)\atop{5.87;4.15}}$
&${0.30;0.46\atop{0.29;0.25}}$
\\
&&& \tiny{$\times 10^{3}$}  &  & \tiny{$\times 10^{-4}$} & \tiny{$\times 10^{-4}$}
& \tiny{$\times 10^{-3}$}& \tiny{$\times 10^{-3}$} 
& \\
\hline
4U 1702-429  & 1.45$^{+0.05}_{-0.05}$ & $12.65^{+0.90}_{-0.80}$&
3.04
& \cite{JN17}
&${0.27;0.32\atop{0.27;0.32}}$
&${0.27;0.32\atop{0.27;0.33}}$
&${0.33;0.38\atop{0.33;0.39}}$
&${3.35;3.00\atop{2.89;2.79}}$
&${0.19;0.18\atop{0.17;0.17}}$
\\
\hline
4U 1608-52  & 1.74$^{+0.14}_{-0.14}$ & $9.3^{+1.0}_{-1.0}$&
1.62
&\shortstack{\cite{GO10}\\\cite{ZR20,RK23,FO16}}
&${0.51;0.47\atop{0.40;0.44}}$
&${0.53;0.47\atop{0.40;0.44}}$
&${-2.53;0.74\atop{0.72;0.57}}$
&${-(1.95;9.27)\atop{-4.60;19.0}}$
&${0.10;0.24\atop{0.102;0.096}}$
\\
\hline
4U 1820-30  & 1.58$^{+0.06}_{-0.06}$ & $9.1^{+0.4}_{-0.4}$&
1.40
& \cite{TG10,TG13,ZR20,FO16}
&${0.47;0.48\atop{0,45;0.47}}$
&${0.45;0.47\atop{0,45;0.47}}$
&${0.64;0.62\atop{0.65;0.63}}$
&${0.64;0.62\atop{0.65;0.63}}$
&${0.102;0.097\atop{0.090;0.090}}$
\\
\hline
J1814-338  & $1.21^{+0.05}_{-0.05}$ & $7^{+0.4}_{-0.4}$&
3.18
& \cite{MB13,YK24}
&${0.16;0.16\atop{0.15;0.16}}$
&${0.16;0.17\atop{0.15;0.16}}$
&${0.28;0.23\atop{0.22;0.21}}$
&${-(6.65;23.2)\atop{-16.6;38.9}}$
&${0.17;0.33\atop{0.27;0.42}}$
\\
\hline
J0437-4715  & 1.45$^{+0.05}_{-0.05}$ & $12.65^{+0.90}_{-0.80}$&
5.747
& \cite{JN17}
&${0.13;0.15\atop{0.13;0.15}}$
&${0.13;0.15\atop{0.13;0.15}}$
&${0.16;0.18\atop{0.16;0.18}}$
&${1.99;1.59\atop{1.68;1.47}}$
&${0.19;0.18\atop{0.18;0.17}}$
\\
\hline
J0614-3329  & 1.45$^{+0.05}_{-0.05}$ & $12.65^{+0.90}_{-0.80}$&
3.15
& \cite{JN17}
&${0.62;0.80\atop{0.34;0.84}}$
&${0.02;0.80\atop{-0.79;0.85}}$
&${0.001;1.02\atop{-0.01;1.10}}$
&${0.11;4.72\atop{-0.28;4.87}}$
&${0.024;0.24\atop{0.040;0.236}}$ 
\\
\hline
J0952–0607A & 2.35$^{+0.17}_{-0.17}$ & $12.245^{+0.685}_{-0.315}$&
1.41
& \cite{RK23}
&${0.73;0.72\atop{0.62;0.65}}$
&${0.75;0.73\atop{0.63;0.65}}$
&${2.60;1.35\atop{0.98;0.91}}$
&${-(4.39;7.61)\atop{-(12.2;66.0}}$
&${0.17;0.23\atop{0.27;0.65}}$
\\
\hline
Vela Pulsar & $1.88^{+0.16}_{-0.16}$ & $12.91^{+0.39}_{-0.39}$&
89.3
& \cite{DS24}
&${0.97;1.02\atop{0.95;1.02}}$
&${0.75;0.73\atop{0.63;0.65}}$
&${2.60;1.35\atop{0.98;0.91}}$
&${9.04;3.14\atop{1.71;1.43}}$
&${0.59;0.35\atop{0.24;0.23}}$
\\
&&&&  & \tiny{$\times 10^{-2}$} & \tiny{$\times 10^{-2}$}
&\tiny{$\times 10^{-2}$}&\tiny{$\times 10^{-1}$}
& 
\\
\hline
Vela X-1& $1.77^{+0.08}_{-0.08}$& $9.56^{+0.08}_{-0.08}$ & $2.83$
& \shortstack{\cite{MR11,TG13}\\\cite{SS23,GN23}}
&  $ {2.62;2.62\atop{2.45;2.47}}$
&  $ {2.66;2.65\atop{2.45;2.47}}$
& $ {4.96;4.67\atop{3.66;3.59}}$
  & $ {4.96;4.67\atop{-(6.52;8.58)}}$
&  $ {0.12;0.12\atop{0.25;0.29}}$
\\
& $2.12^{+0.16}_{-0.16}$& & &
&  $ {3.88;3.75\atop{2.81;2.79}}$
&  $ {4.13;3.94\atop{2.84;2.82}}$
& $ {-(3.49;5.01)\atop{7.23;6.33}}$
& $ {-(1.02;1.10)\atop{-(1.94;2.17)}}$
& $ {0.10;0.11\atop{0.14;0.15}}$
\\
&&& \tiny{$\times 10^{5}$}& & \tiny{$\times 10^{-6}$} & \tiny{$\times 10^{-6}$}
&\tiny{$\times 10^{-5}$}&\tiny{$\times 10^{-5}$}
& \\
\hline
${SAX~J1748.9\atop{-2021}}$ & 1.45$^{+0.05}_{-0.05}$ & $12.65^{+0.90}_{-0.80}$&
3.04
& \cite{JN17}
&${0.37;0.41\atop{0.30;0.43}}$
&${0.38;0.41\atop{0.30;0.43}}$
&${0.97;0.53\atop{0.38;0.51}}$
&${-2.61;15.2\atop{6.34;3.86}}$
&${0.15;0.39\atop{0.21;0.17}}$
\\
\hline
EXO 1745-268  & 1.45$^{+0.05}_{-0.05}$ & $12.65^{+0.90}_{-0.80}$&
3.04
& \cite{JN17}
&${0.46;0.50\atop{0.37;0.52}}$
&${0.47;0.51\atop{0.37;0.52}}$
&${1.41;0.66\atop{0.48;0.63}}$
&${-2.95;18.8\atop{10.9;4.94}}$
&${0.13;0.35\atop{0.23;0.16}}$
\\
\hline
KS 1731-260  & 1.45$^{+0.05}_{-0.05}$ & $12.65^{+0.90}_{-0.80}$&
1.91
& \cite{JN17}
&${0.52;0.45\atop{0.29;0.48}}$
&${0.57;0.45\atop{0.29;0.48}}$
&${-0.29;0.59\atop{0.38;0.088}}$
&${-1.20;26.8\atop{12.1;3.97}}$
&${0.085;0.46\atop{0.25;0.15}}$
\\
\hline
\end{tabular}

\vspace{-0.2cm}
\caption{{
 Observable 
        data 
    for masses $M$ (second), radii $R$ (third) up to J0614-3329,
    rotational periods $P$ (fourth), references (fifth). For other
        NSs the radii $R$ are found by using the model calculations.
    The calculated NS angular momentum
    $\overline{\omega}=Ic/(M^2G)$, Eq.~(\ref{rotpar}), for $f/R=0.03$,
    $\kappa=10$ ($q=0$ for the separation constant,
    the spherical case), $0.1$,$10$ ($q=1$),
    $0.3$,$10$ ($q=1$), and
    $0.1$, $50$ ($q=1$) in sixth, seventh, eigth, and ninth column,
    respectively.
   The upper evaluation for NS
       periods $P_0$ (its absolute values), Eq.~(\ref{om0}),
       for the adiabatic condition,
       Eq.~(\ref{omcond}), for  $f/R=0.1$, $\kappa=50$ ($q=1$) in
     tenth column
     for
     the NSs named in the column 1.
        The relative
   crust thickness is given by
   $a/R=0.08$.
     Lower and upper rows 
        in
        each NS line 
        are related to the
        minimal and maximal values of the NS mass $M$, according to the
        errors
        in the second column.
        Minimal (left) and maximal (right) values of the NS radii $R$,
        separated by
        a semicolon, correspond to their mean values with
        errors shown in the third column.
      }}    
\label{table-1}
\end{table*}

\section{Discussions of the results}
\l{disc}
\begin{figure}[htp]
  \begin{center}
    \includegraphics[width=0.5\textwidth,clip]{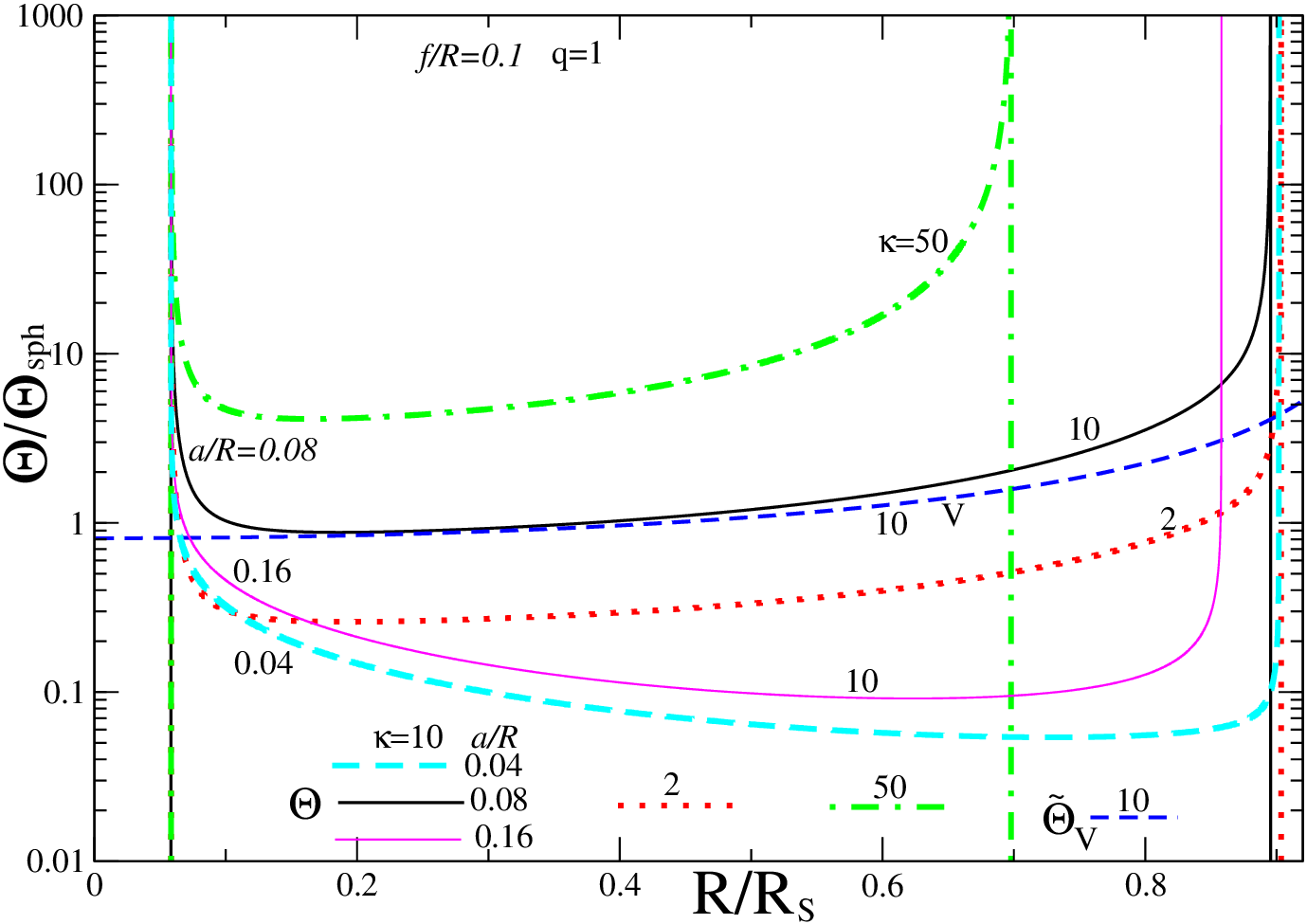}
    \end{center}

  \caption{{\small The same as in Fig.~\ref{fig3} but for different nuclear
      gravity incompressibilities 
      $\kappa=2$ (weak, red dotted), $10$ (strong, black solid),
      $50$ (super strong gravity,green dashed),
      Eq.~(\ref{kappa}) ($a/R=0.08$), and for a few more examples
      of a small leptodermic
      parameter $a/R$ (cyan dashed $a/R=0.04$ and
      magenta solid, $a/R=0.16$ for $\kappa=10$).
      The blue dashed line displays
      the statistically averaged volume contribution at $f/R=0.1$, and
      $\kappa=10$.
   }}
\label{fig4}
\end{figure}

\begin{figure}[htp]
\begin{center}
       \includegraphics[
    width=0.5\textwidth,clip]{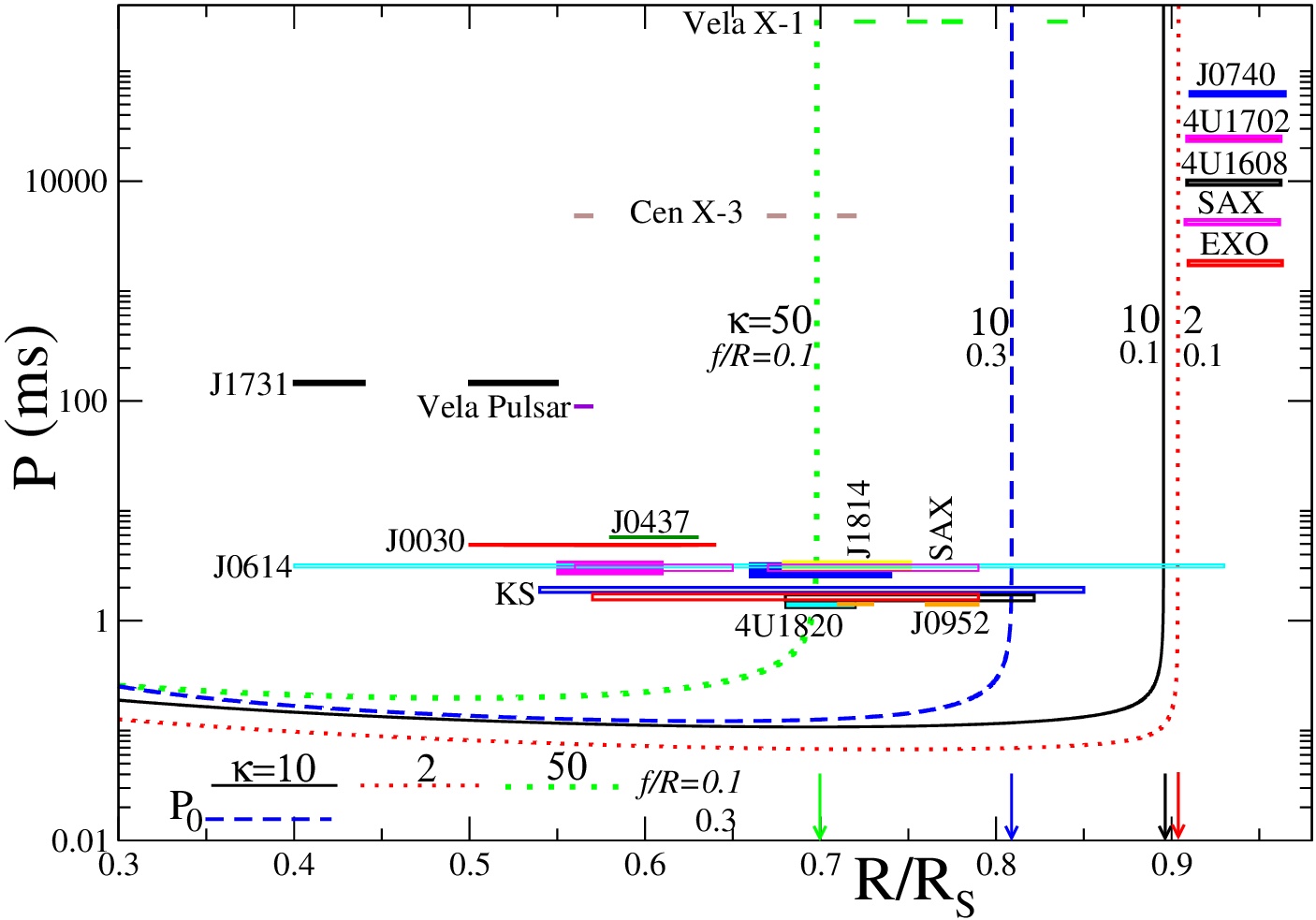}
       \end{center}
  
  \caption{{\small   NS periods $P$ (in units of ms) are plotted as functions
        of dimensionless radius variable $R/R_{\rm S}$. 
            Straight line segments
        with different colors and boundaries
        show different 
        data for NSs presented within the intervals of
        $R/R_{\rm S}\approx \sqrt{2MG/R c^2}$;
        see Eq.~(\ref{boundcond1}) and Table \ref{table-1},
        second and third columns). 
             Different lines are
        the characteristic
         periods $P_0$, Eq.~(\ref{omcond}), for 
        the full
        gravitational  metric with
        accounting for the $t,\varphi$ correlations and surface
        contributions for the deformed $q=1$ case
        at the relative
        incompressibility values 
        $\kappa$=2,  
        10. and $50$;
        see Eq.~(\ref{kappa}) for $\kappa$. 
        The leptodermic and deformation parameters are 
        given by
        $a/R=0.08$ and $f/R=0.1$ (green dotted) and $f/R=0.1$
        (blue dashed lines). NS radius is fixed by the averaged value,
        $R=12$ km.
             Arrows show the
        corresponding asymptotes.
   }}
\label{fig5}
\end{figure}

\begin{figure*}[htp]
  \begin{center}
  \includegraphics[clip,width=5.8cm]{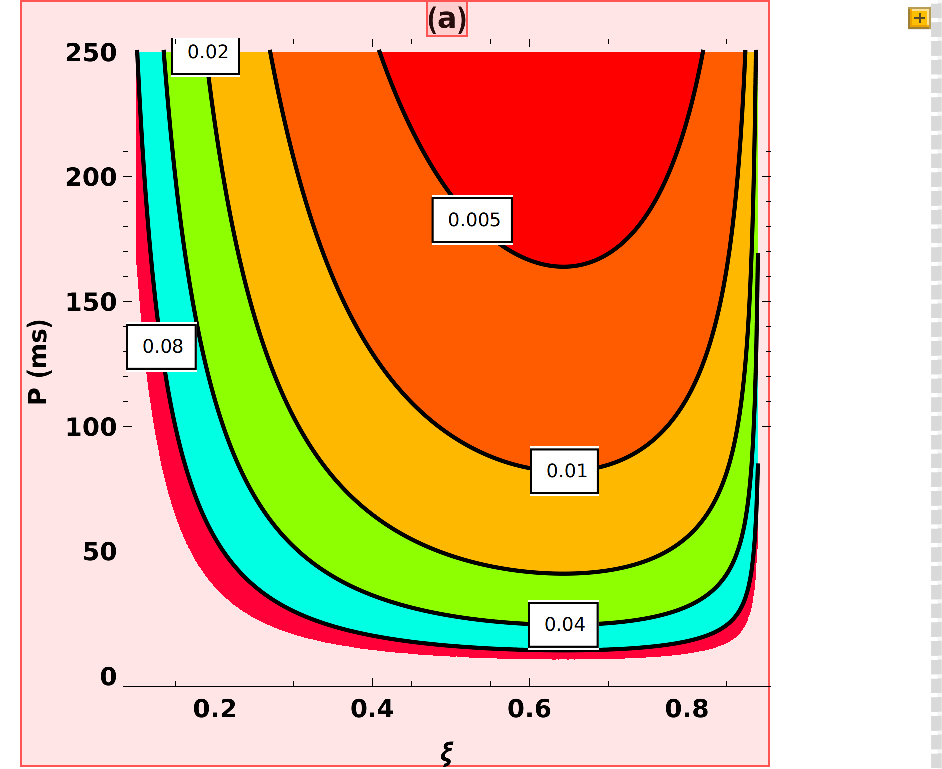}
  ~
  \includegraphics[clip,width=5.8cm]{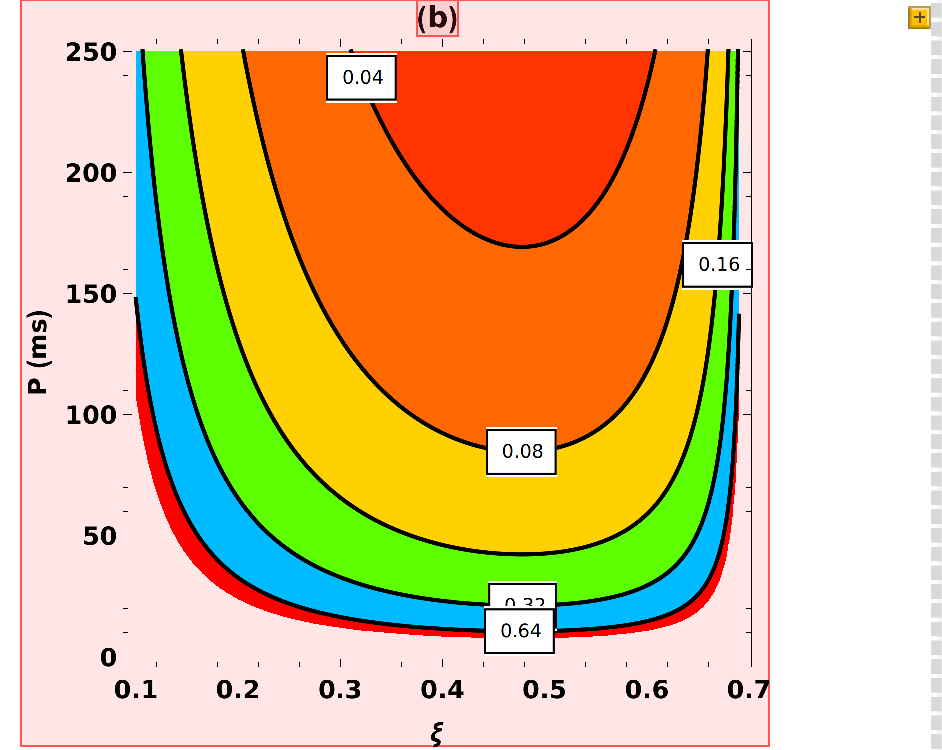}
  ~
  \includegraphics[clip,width=5.8cm]{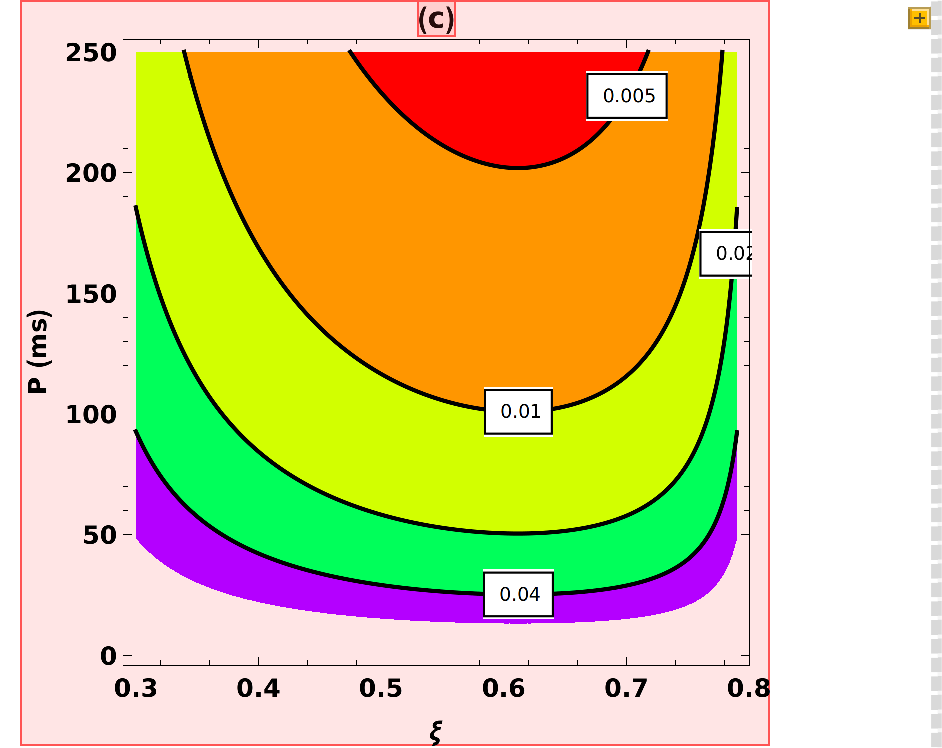}
  ~
  \includegraphics[clip,width=5.8cm]{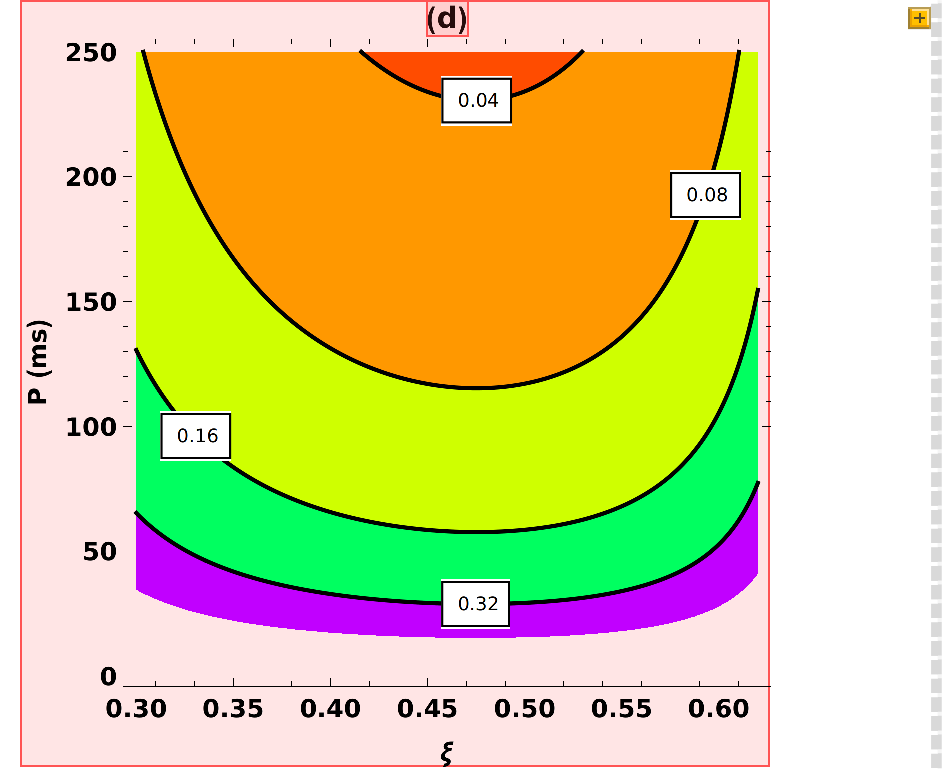}
  \end{center}

  \caption{{\small
      Contour plots show the 
      NS angular momentum parameter,
      $Ic/(M^2G)\approx 2\pi \Theta c/(M^2 P G)$, Eq.~(\ref{rotpar}),
      as function of the dimensionless variable
      $\xi=R/R_{\rm S}$ and period (in milliseconds) for the
      metric coefficient
      $\tau(r,\theta)$, Eq.~(\ref{14tauinexpsol}).
      The typical values of $Ic/(M^2G)$
      are shown in squares. (a): For incompressibility $\kappa=10$
      at the deformation $f/R=0.1$;
      (b): $\kappa=50$, $f/R=0.1$ . 
      (c): $\kappa=10$, $f/R=0.3$
      (d):  $\kappa=50$, $f/R=0.3$.
          A mean radius is $R=12$ km, and a crust thickness $a/R=0.08$,
           as in Fig.~\ref{fig5}.
           White areas are related to non-physical
           parameters values.
          }}
\label{fig6}
\end{figure*}

Figures \ref{fig3} - \ref{fig6} and Table \ref{table-1}
show our results. The moments of inertia $\Theta$,
Eq.~(\ref{MI}), as
functions of the NS radius variable  $R$ in units of the Schwarzschild radius
$R_{\rm S}$, Eq.~(\ref{RS}), 
are displayed in Figs.~ \ref{fig3} and  \ref{fig4}
for different values of parameters.
Figure \ref{fig3} presents the MI  depending on small surface deformations $f/R$
(or the spheroid deformation $Q$).
As seen clearly from this figure, the most remarkable feature of the MI is
existence of the minimum
between the two asymptotes.These asymptotes are related to the poles of the MI,
Eq.~(\ref{MI}), obtained because of the approximate self-consistency
as shown in Appendix \ref{appB}. Therefore, one appears the NS radius
constraints due to the rotation as in the spherical case
\cite{AM25npae,MM26npa}. However, with a surface deformation,
such constraints restrict the allowed radii
from both sides.
With increasing deformations
$f/R$ the allowed radii range becomes significantly smaller. Notice that
the numerical ``num'' calculations of the correlation contribution
$\mathcal{T}_{t\varphi}$, Eq.~(\ref{MIeq6appF}) (see the black solid curve), is
mainly in good agreement with the analytical ``an'' result,
Eq.~(\ref{MIeq7appF})
(red-dotted curve). For other contributions we use the analytical
expressions; see discussions in Section \ref{SFA}, especially 
Fig.~\ref{fig2}.  An exclusion can be done for small ranges near
the asymptotes where the surface components should be calculated more
accurately \cite{MM26npa}. Notice that we found close results for the
separation constants $q=0$ and $1$~.

Figure \ref{fig4} displays the effects of the analytical MI expressions
on the effective
nuclear-gravitational incompressibility
$\kappa$, Eq.~(\ref{kappa}), and the relative crust thickness $a/R$.
The upper asymptote is
decreasing and the MI values are
increasing much with the incompressibility $\kappa$ from $\kappa=2$ (weak)
to $\kappa=10$ (strong), and  $\kappa=50$ (super strong gravity).
Therefore, the rotation
constraints areas are essentially shortened with $\kappa$. A similar behavior
of the upper asymptote is seen for increasing the crust
thickness $a/R$ but much weaker within its small values. More dramatic
crust thickness effect can be noted near the lower asymptote position
which is, however, the same
for all presented curves; cf. dashed cyan ($a/R=0.04$), black  solid
($a/R=0.08$), and thin magenta solid  ($a/R=0.16$).
All of them have qualitatively similar minima
between these two asymptotes. These surface and correlation
pecularities very much differ from
the smooth statistically averaged volume MI contribution,
$\tilde{\Theta}_V$ (the blue dash line). 
Notice that the lower constraint is usually
much larger than the
gravitational radius $r_g$, Eq.~(\ref{rg}), which determines the events
horisont \cite{LLv2} in the non-rotating systems. According to the
boundary condition, one has $r_g/R_{\rm S}\approx (R/R_{\rm S})^2$, and
this variable is much smaller than all MI poles for $R/R_{\rm S}\siml 0.9$.

Table \ref{table-1} shows our results
for the dimensionless angular momentum, $Ic/M^2G$,
Eq.~(\ref{rotpar}), 
    for different deformations $f/R$ and incompressibilities $\kappa$.
Upper
period limits $P_0$, Eq.~(\ref{om0}), are shown for extremal parameters
$f/R=0.3$ and incompressibility $\kappa=50$.  
    Both criteria, $Ic/M^2G\ll 1$ and $P_0\ll P$ [Eq.~(\ref{omcond})],
    are important for checking macroscopic
adiabatic conditions. The observable data for the both NS masses $M$
and the radii $R$ are displayed above the NS J0952–0607A line, and others for
the calculated radii $R$ using NS models.
   Analytical calculations are used for the
   MI $\Theta$, Eqs.~(\ref{MI}) [with (\ref{MIt})
     and (\ref{MItf}), (\ref{adMIVS}), (\ref{MIeq1appF}),
     (\ref{MIeq4appF}), (\ref{MIeq7appF}),
     (\ref{adMItS}), and (\ref{adMItfS})]; see also Eq.~(\ref{resc1})
     for the constant $c^{}_1$. For a small deformation,
     $f/R=0.1$, and strong gravity incompressibility $\kappa=10$,
     the adiabatic condition, Eq.~(\ref{omcond}), in terms of (the
     absolute value) of $P_0$ 
     is carried out well
     for all NSs,
     that is not shown in Table. For $\kappa=50$ the parameter $P_0$
     (last column) is good everywhere above the J0952 NS line.
     Negative values of $Ic/M^2G$ occur for the cases when
     the data are out of the physical range between the roots of the
     denominator in Eq.~(\ref{MI}); see also the last column. 
     The dimensionless angular momenta, $Ic/M^2G$,
     are mainly close for the deformed $q=1$ and the
     spherical $q=0$ cases. For a strong incompressibility? $\kappa=10$,
     they are mostly  
     small for all the NSs, except for
     4U 1608-52,~4U 1820-30,~J0614-3329,~J0952–0607A, and three last NSs.
     Negative values of
     $Ic/M^2G$ (see 4U 1608-52,~J0614-3329 and last NS) mean that the
     data occur in a non-physical
range which is out of the
rotational constraints range within the first order perturbation approach;
see below in more details.
For the intermideate
situation in the NSs J0030+0451,
~J0740+6620, ~4U 1702-429, ~4U 1820-30, and SAX J1748.9-2021, the quadratic
order for the parameter $Ic/M^2G$  ($\overline{\omega}^2$) are nevertheless
sufficiently
small such that they can be largely neglected.
For larger deformations ($f/R=0.3$, eigth column), the situation
concerning the slow rotational conditions becomes
worse, especially with increasing
the incompressibility $\kappa$ up to $50$ (seventh column).  Even at this
super strong gravity $\kappa=50$, the rotational parameter $Ic/M^2G$
for NSs J1731-347, Cen X-3,
Vela pulsar and Vela X-1 are remaining small. Notice that for the strong gravity
$\kappa=10$, the NSs above the 4U 1608-52 line and for the
 ~J1814-338 and
J0437-4715 NSs, one can largely neglect the terms with a small
$Ic/M^2G$, and sometimes, those of the order of 
$(Ic/M^2G)^2$.
Notice that
for the adiabatic condition, Eq.~(\ref{omcond}), in terms
(of the absolute values)
of $P_0$
(see the ninth column)
in the specific deformed-surface case ($f/RT=0.3, q=1$) at a supper gravity
incompressibility
($\kappa=50$), is carried out formally well, except for the NS SAX J1748.9
    -2021 at a smaller radius $R$ and a smaller
     mass $M$ boundary. This takes place though  the rotational parameter 
     $Ic/M^2G$ for the most of discussed NSs in this extremal situation
     occurs non-physically negative out of constraints;
     as mentioned above.

Figure~\ref{fig5} displays
      the characteristic limit periods  
      $P_0$,  Eq.~(\ref{omcond}),  
      as function of the ratio of the
      effective radius $R$ 
      to  the Schwarzschild radius 
      $R_{\rm S}$, Eq.~(\ref{RS}),  $R/R_{\rm S}$,
      versus the results calculated   
      for the data on the mass $M$, radius $R$
      and period $P$ for all NSs in
      Table \ref{table-1}. We present
      them at
      the definite periods $P$
         because of good accuracy of these 
         measurements,
             besides of the NS EXO J1745-268 where we neglect
             small period errors for simplicity.
The intervals of the variable $R/R_{\rm S}$
         are presented by line segments 
         within the 
             errors for the mass $M$ (second) and radii $R$ (third 
                 columns) in
         Table \ref{table-1}. 
         For the calculation of $P_0(R/R_{\rm S})$,
         this variable is expressed 
       through the mass $M$ 
       and radius $R$ by using both the  
       boundary outer-inner slitching condition,
       Eq.~(\ref{boundcond1}),
        and the gravitational radius 
           $r_g$ value, Eq.~(\ref{rg}),
         $R/R_{\rm S}=[2GM/(Rc^2)]^{1/2}$.
    The radius $R$ is taken 
        approximately
  as a mean 
    radius for NSs in Table \ref{table-1}, $R=12$ km.
    As the MIs $\Theta$, the limit periods $P_0$, Eq.~(\ref{omcond}),
    are dramatic 
       functions
      of the relative effective NS radius, 
      $R/R_{\rm S}$, with the
      corresponding asymptotes 
      $R=R_{\rm K}$, 
      similarly as in Figs.~\ref{fig3} and \ref{fig4}. 
      The critical period functions 
      $P_0(R/R_{\rm S})$ for $\kappa=2~-~50$ having a minimum between
      the two asymptotes are 
                    monotonically increasing, being 
      convergent to the corresponding 
      asymptotes $R=R_{\rm K}$. 
     In Fig.~\ref{fig5} we show also the essential dependence  of
      the full limit periods $P_0$
      on the dimensionless incompressibility parameter $\kappa$
      [Eq.~(\ref{kappa})]
  through the tension coefficient $\sigma$, 
 Eq.~(\ref{enertots}),
 in the surface MI and energy contributions, and on the surface deformation
 $f/R$.
 The values of $P_0$
 increases with $\kappa$ from a weak ($\kappa=2$) to
 strong ($\kappa=10$), anf then, super strong ($\kappa=50$)
 gravitation for the radius $R$ within the rotational constraints (MI poles)
 $R_{\rm K}/R_{\rm S}$ in Eq.~(\ref{MI}).
 As a result of these analytical calculations,
 for a strong $\kappa=10$ (solid black)
the adiabatic condition, Eq.~(\ref{omcond}), is carried out well
 for
 the most of neutron stars
 shown in Table \ref{table-1}, 
 except for those having the negative MI or  $Ic/(M^2G)$. For stronger gravity
 ($\kappa=50$) and larger deformation ($f/R$) this condition is
 fulfilled worse for most of NSs, in contrast to the spherical case
 \cite{MM26npa}.
       The surface effect is 
      enhanced because of the correlation contribution
      $\mathcal{T}_{t\varphi}$ 
      due to a change of the rotational root
      $R=R_{\rm K}$. 

      For the binary system GW170817, for which the individual
      NS radii $R$ and
      masses $M$ are obtained simultaneously with especially good precision
      \cite{BA19-1}, the
      rotational NS periods cannot be measured  with  a good accuracy but
      the dimensionless angular momenta $\overline{\omega}=Ic/M^2G$
      can be estimated.
      In the Low-spin-prior case, $\overline{\omega}\leq 0.04$,
      one approximately finds from our calculations the upper
      constraints $\overline{\omega}^{}_1=0.037-0.038$,
      $\overline{\omega}^{}_2=0.037-0.04$ for periods $P_1=P_2=27$ ms while
      for the High-spin-prior case $\overline{\omega}\leq 0.50$ for larger mass
      $M_1$
      and $\leq 0.61$ for $M_2$, one obtains
      $\overline{\omega}^{}_1=0.45-0.47$ and
      $\overline{\omega}^{}_2=0.53-0.57$ for the periods
      about $2.2$ ms. For the first case, one can apply the
      first-order  perturbation approach for NSs GW170817, in contrast
      to the opposite case.
     
       Thus, as clearly seen from Fig.~\ref{fig5} 
      and Table \ref{table-1},
      the macroscopic adiabaticity condition (\ref{omcond}) 
      is carried out well
      for almost all NSs with observed masses and radii,  at 
      strong gravity  $\kappa=10$, even for larger surface deformation
      $f/R$. 
          The comparison becomes 
       worse with increasing gravitation, e.g., for $\kappa=50$.
             In line of the results for the dimensionless
       angular momentum $Ic/(M^2G)$ (Table \ref{table-1}),
       it would be nice to take into account 
       non-adiabatic effects for several NSs. They are related, e.g.,
       to the quadratic frequency corrections
       $\sim \overline{\omega}^2$,  
           especially for
           the rotation of neutron stars 
               4U 1608-52, ~4U 1820-30, and J01614
             with observable masses and radii, and perhaps for
           almost all NSs with model-evaluated radii and extremaly short
           periods.
                  In this non-adiabatic approach,
                  the NS MI $\Theta(\omega)$ depends essentially on the
                  rotation
                  frequency $\omega$. 

    Contour plots in Fig.~\ref{fig6} 
       show the analytical calculations of the rotational
parameter, $\overline{\omega}(R/R_{\rm S},P)$, Eq. (\ref{rotpar}), depending on
the dimensionless variable $\xi=R/R_{\rm S}$ and rotational
period $P$.  The intervals mostly
overlap data presented in Table \ref{table-1},
except for Cen X-3 and Vela X-1 with extremally large periods. As seen
from all contour plots (a-d) for the deformations $f/R=0.1$ (a,b) and
$0.3$ (c,d)
and incompressibilitieas $\kappa=10$ (a,c)
and  $\kappa=50$ (b,d), the parameter $\overline{\omega}$ has a
minimum but in different areas of the variable $\xi$. With increasing
incompressibilities $\kappa$, the rotational parameter values $Ic/M^2G$ are
increased
for any deformation, and the allowed $\xi$ areas for their definition
become smaller. With increasing deformation $f/R$, these minima
slightly decreased. In both cases, the non-physical areas become
larger for smaller
periods, in line of the results presented in Table \ref{table-1}.   
 The rotation parameter $\overline{\omega}$,
Eq.~(\ref{rotpar}),  has a minimum because 
of the maxima of the NS mass $M$,
and minima of  the MI
$\Theta$ within the allowed physical values between the asymptotes
at
the 
rotational MI root $R_{\rm K}$. As seen from all
these contour plots, the values of 
$\overline{\omega}$ are mainly small,
    except for the areas of very small
    periods and very large $\xi$. They become larger with increasing both the
  incompressibilities $\kappa$ and deformations $f/R$. 

  In line of all these discussions
     we may conclude that the
     linear perturbation approximation fails in the case of
     $\overline{\omega} \simg 1$ (or negative)
         for several neutron stars with their very short periods.
     Therefore, in  this case we
     should study
     a non-linear (non-adiabatic) approach, taking into account, e.g., the
     centrifugal $\overline{\omega}^2$ corrections, that might be a
     challenge for
     a forthcoming work.
 
     \section{Conclusions}
\l{concl}

     The macroscopic effective-surface approximation based on the leptodermic expansion
     over a small parameter - a ratio of the crust thickness $a$
     to the effective radius $R$ of the system - is extended for
     the description
     of neutron star rotations by taking into account the
     surface deformations.
   The neutron star was considered  as 
   a finite perfect dense-liquid (including amorphous solid)
   system rotating stably with a small constant frequency $\omega$
   around a symmetry axis within the Kerr approach by using the
   spherical coordinates. 
   Within the linear approximation in $\omega$ around the Schwarzschild
   spherically symmetric system, one can derive the
   Cristoffel Symbols and Ricci tensor for the GRT equations, which are linear
   in the Riemann tensor, for
   the radial variable $r$ larger, $r>R$, and smaller, $r<R$,
   regions than the mean finite effective radius $R$ for different values
   of the integration separating constant $q$.
     In this approximation we show that the
     outer solution ($r>R$) for the off-diagonal
     gravitational metric element $g^{}_{03}$ is
     the Boyer-Lindquist
     result for the spherical case $q=0$, obtained earlier
     for the vacuum (or for a
     point gravitational system). But it differs essentially for
     the deformed NS surface at
     $q \neq 0$. For $q<9/4$ our outer solution for $\tau(r,\theta)$
     coefficient in $g^{}_{03}$  tends to zero for large $r$ limit.
        This result for the outer off-diagonal
     $\alpha^{}_K$ dependent term for a small deformations 
     of a neutron star with a finite size $R$
     is expressed in terms of the Schwarzschild metric functions
        in our
     linear perturbation approach, $\overline{\omega}\ll 1$.
     In particular, one can use the
     Equation of State (EoS) with the macroscopic
     energy density, $\epsi=\epsi(\rho)$ as functional of the density
    $\rho$, in a rather general form
     taking into account the gradient surface terms.
      We derived also the rotational off-diagonal gravitational metric component
     $g^{}_{03}$ [the coefficient $\tau(r,\theta)$]
     for the inner NS region, $r<R$,
     in the same linear approximation, $\overline{\omega}\ll 1$, for small NS
     radius $R$ with respect to the larger Schwarzshild radius,
     $R_{\rm S}\propto
     1/(\overline{\rho})^{1/2}$,
     where $\overline{\rho}$ is the inner mass density.
     The integration constants of $\tau$ were found from the boundary
     conditions, - zero in the limits $r \rightarrow 0$ and $\infty$,
     and finite matching of the inner with outer solutions in the laptodermic
     approximation valid for any finite dense matter system.
        With the help of our GRT solutions
     we obtained the NS adiabatic MI $\Theta$ in terms of the
     statistically averaged  $\tilde{\Theta}$ and rotational correlational
     $\mathcal{T}_{t\varphi}$ contributions. Each of them are presented
     as a sum of the
     volume and surface terms.  These MIs of the NSs with the deformed
     effective surface
     change dramatically as functions
     of the dimensionless angular momentum $Ic/M^2 G$ (through 
     the off-diagonal $g^{}_{03}$ gravitational-metric  component),
     the
     effective radius $R$, and dimensionless
     incompressibility $\kappa$, due to the
     $t,\varphi$ correlational 
     contribution $\mathcal{T}_{t\varphi}$.
     The significant macroscopic surface effects
     were also found and studied in relation to this correlation term
     for a strong gravitation.

  As perspectives, one can generalize  our analytical approach to 
    the many-component deformed rotational systems. 
    taking into account  the symmetry energy
  of the isotopical asymmetry 
  within
      the Kerr metric approach at small rotational frequencies and
      deformations.
             This peculiarity should be taken into account 
         going beyond the linear perturbation approach for larger
         rotational frequencies of the neutron stars.
 We emphasize
 that the ESA theory \cite{MM24}
 allows to consider the deformed and super-deformed 
     systems like those
 in the rotating tidal accretion binary NSs.
                We are planning also to derive the corrections
         to the TOV equations for rotational systems, taking into account the
         $t,\varphi$ correlations.

         \acknowledgments

The authors greatly acknowledge  
O.Y.\ Dzyublik, V.Z.\ Goldberg,
M.I.\ Gorenstein, V.P.\ Gusynin, F.A.\ Ivanyuk,   
J.\ Holt, C.M.\ Ko, E.I.\ Koshchiy, J.B.\ Natowitz,
S.A.\ Omelchenko, A.I.\ Sanzhur,
G.V.\ Rogachev, Yu.V.~Shtanov, B.I.\ Sidorenko, 
Yu.V.~Taistra, S.I.~Vacaru, V.I.~Zhdanov
for many creative and useful discussions.

\appendix

\setcounter{equation}{0}
\renewcommand{\theequation}{A\arabic{equation}}

\section{NS macroscopic energies and masses}
\label{appA}

The total ETF macroscopic energy, $E$, 
for zero rotational frequency $\omega$ can be written in terms
of the EoS energy density $\epsi(\rho)$, Eq.~(\ref{enerden}), 
similarly as in Refs.~\cite{strmagden,BM15},
\be\l{energytot}
E=E_{\rm ETF}=\int \epsi \left(\rho({\bf r})\right)\;\d \mathcal{V}.
\ee
%
In our macroscopic derivations up to 
quadratic terms of the order 
of $\overline{\omega}^2$ in the angular momentum $I$, 
it is convenient to introduce also the mass density
$\rho$ as a mass of a given 
small volume of the stellar matter per unit volume.
 Formally, one can connect this mass density $\rho$
with the particle number density $n$ introduced above
by $\rho=m n$, where $m$ is the same test-particle 
mass
as in Sect.~\ref{GRT} (see Ref.~\cite{MM26npa}).
  This test particle is considered as free
in an arbitrary gravitational field \cite{LLv2}. 
In particular, one can introduce 
a more general inner asymptote of the mass density, 
$\overline{\rho}= m \overline{n}$, 
    where $\overline{n}$ is the mean
particle-number density. The corresponding 
Schwarzschild radius $R_{\rm S}$,
 Eq.~(\ref{RS}), can be expressed \cite{MM24,MM26npa} in terms of such 
a density $\overline{\rho}$ through
$\epsi_0$ [Eq.~(\ref{epsi0})]
(Tolman directly considered $\overline{\rho}$ 
instead of $\epsi_0$
in the definition of $R_{\rm S}$, Eq.~(\ref{RS}), in natural units
of $c=G=1$ in 
Ref.~\cite{RT87}).
The gravitational field is approximately included into 
the definition of the mass
density as accounting for the defect mass \cite{LLv2}
 in our macroscopic approach (similarly to the effective mass
 in nuclear physics) through the energy density component $\mathcal{U}$;
 see Sect. \ref{GRT}.
 
 For the macroscopic NS mass $M$
 in the simplest 
case of the interaction, $\mathcal{B}\approx \mathcal{C}$, 
one obtains
\cite{MM26npa}
\be\l{MNStot}
M=\int \rho \d \mathcal{V} \approx  M_{V} + M_{S}~.
\ee
Here, $M_{V}$ is the volume component, 
\be\l{MNSV}
M_{V}
=
2\pi \overline{\rho} R_{\rm S}^3f\left(R/R_{\rm S}\right)~,
~~~f(\xi)=\arcsin(\xi) - \xi \sqrt{1-\xi^2}~,
\ee
with the Schwarzschild radius $R_{\rm S}$ given by Eq.~(\ref{RS}), 
%
%
for $0 < \xi=R/R_{\rm S}< 1$. 
In Eq.~(\ref{MNStot}), $M_{S}$ is the
surface component of the NS mass,
\be\l{MNSSfin}
M_{S}\approx -8 \pi 
\overline{\rho} R^2 a \left(1-R^2/R^2_{S}\right)^{-1/2}, 
\ee
where $a$ is the NS inner crust thickness, $a\ll R$ in the
leptodermic approximation.
Similarly,
we may present the total ETF energy $E$,
 Eq.~(\ref{energytot}), as in Ref.~\cite{MM24},
\be\l{enertot}
E \approx E_V+E_S~,
\ee
where
$E_V$ is the volume part of the total energy,
\be\l{enertotv}
E^{}_V
\!=\!
2 \pi \overline{\rho} c^2
R_{\rm S}^3 f\left(R/R_{\rm S}\right);
\ee
see Eq.~(\ref{MNSV}) for $f(\xi)$.
In Eq.~(\ref{enertot}), 
for the
surface part one obtains
\be\l{enertots}
E_{S} \approx \sigma S~,\qquad 
\sigma=\frac{4a \overline{\rho} K_G}{135 m} J(R),
\ee
 where $S=4\pi R^2$ is the surface area value,  
 and $\sigma$
 is the leading-order tension coefficient \cite{MM24,AM25npae,MM26npa}. 
The gravitation is taken into account through the
coefficient $J(R)$, which is the Jacobian $J(r)$,
Eqs.~(\ref{Jin}) and (\ref{Jout}),
at the effective surface $r=R$.
Thus,
one obtains the explicit analytical expression for the
tension coefficient $\sigma$, proportional to the 
leptodermic parameter,
\be\l{leppar}
a/R = (18\mathcal{C} \overline{\rho}/K_G)^{1/2}~,
\ee
where $\overline{\rho}$ is
the same inner density and  $K_G$ is the
incompressibility $K_G$, Eq.~(\ref{KG});
see, again, section \ref{GRT}.
    Notice that we neglected  the contribution $\sigma_{\rm corr}$
    to the tension coefficient $\sigma$  [see Eq.~(\ref{enertots})
      and Ref.~\cite{MM26npa}],  
    \be\l{sigmacorr}
    \sigma_{\rm corr}=
    \frac12 a\overline{\rho}  J(R)~,~~~\frac{\sigma_{\rm corr}}{\sigma}
  \approx \frac{135}{96\kappa}
      \approx \frac{1}{\kappa}~.
    \ee
    %
    %
    %
      We introduced here the relative 
      total dimensionless
      incompressibility $\kappa$ (Ref.~\cite{MM24}),
      which is useful in these derivations. The $\kappa$ is
      a measure of the gravitational
      field contribution, in addition
      to the well known nuclear matter value,
    \be\l{kappa}
\kappa=\frac{\overline{\rho} K_G}{12 m\epsi_0}~. 
    \ee
    Here, $\epsi_0$ is determined by Eq.~(\ref{epsi0}) and $K_G$
    is given by Eq.~(\ref{KG}). For nuclear physics,
    one finds $\kappa\approx 1$ ($K_G=K\approx 240 $ MeV,
     $\epsi_0=\rho^{}_0 c^2=-b^{}_V n^{}_0$, $b_V \approx -16$ MeV,
    $n^{}_0=0.16$ fm$^{-3}$). Therefore, for a low gravity, one can
    approximately put
    $\kappa=2$.
  For
    a strong NS gravitational field, 
    one may assume $\kappa \gg 1$. Then, 
    according to the condition Eq.~(\ref{sigmacorr}),
     one can neglect
    the correction $\sigma_{\rm corr}$ 
    to the tension coefficient $\sigma$.

    In order to show importance of the gravitation and surface contributions
in our macroscopic calculations,
one can directly evaluate the energy density $\overline{\epsi}$,
Eq.~(\ref{epsi0}),
    by using the coordinate transformation and boundary condition for 
    slitching together the outer,
$e^{-\lambda}=1-r_g/r$ [see Eq.~(\ref{SchwarzOUT})], and inner,
$e^{-\lambda}=1-r^2/R_{\rm S}^2$, 
Schwarzschild metric [see Eq.~(\ref{SchwarzIN})], at the ES, $r=R$.
Thus, approximately, in the leading order over the leptodermic parameter,
$a/R$, one has Eq.~(\ref{boundcond1}).
Substituting Eqs.~(\ref{RS}) for the Schwarzschild radius $R_{\rm S}$ and
(\ref{rg}) for the gravitational radius $r_g$ to this 
expression for $R_{\rm S}$,
one finds approximately the inner energy, $\overline{\epsi}$, and
density, $\overline{\rho}$,
\be\l{Eps0MR}
\overline{\epsi}= \overline{\rho} c^2 = \frac{3Mc^2}{4\pi R^3}~,\quad
\overline{\rho} \approx
      \frac{3M}{4\pi R^3}~.
      \ee

\setcounter{equation}{0}
\renewcommand{\theequation}{B\arabic{equation}}

\section{Spin of neutron stars}
\label{appB}

For the angular momentum $I$ in the case of the stationary rotation
of a finite perfect fluid around its symmetry axis,
one has the expression
\cite{FIP86,NS03,NS17,AW08,LLv2,LLv6} (\ref{Igen}).
The
gravitational metric $g_{\mu\nu}$ for a small rotational frequency 
parameter, $\alpha^{}_K \propto \overline{\omega}$ (Sect. \ref{GRT}), 
up to $\overline{\omega}^2$ terms, can be
 presented
as done \cite{BL67,PH76,AW08,MV08} in Eq.~(\ref{AWmetgen}).
 In Eq.~(\ref{AWmetgen}),
up to $\overline{\omega}^2\propto \omega^2$ terms, we 
introduced the function $\tau$ found in the Section
    \ref{DRNSM}.
We may compare
this  $\tau$ with that  obtained in
 Ref.~\cite{PH76}; see also Refs.~\cite{MM26npa,AM25npae}.
Under these definitions and assumptions, Eq.~(\ref{Igen})
reduces in the canonical ensemble to
    (see Refs.~\cite{AW08})
\be\l{Igen1}
I=\frac{1}{c^2}\int \epsi(\rho) (u^t)^2
\left(\omega~g_{\varphi\varphi} +
g_{t\varphi}\right)\d \mathcal{V}_{\omega}~.
\ee
The quantity in the integrand parentheses is of 
the linear order over $\omega$ 
    ($g_{t\varphi}\propto \tau \propto \omega$).
Therefore, neglecting terms of the order 
of $\overline{\omega}^2$ to be accurate at linear order over
$\overline{\omega}$, 
one can use the approximation
$\d \mathcal{V}_{\omega}\approx \d \mathcal{V}$, Eq.~(\ref{dV}).
 Neglecting also the Lorentz shortening factor
in the mean  four-velocity, one has
\be\l{ut}
u^t=\frac{e^{-\nu/2}}{(1-v^2/c^2)^{1/2}}\approx e^{-\nu/2}~,
\ee
where $v$ is the test particle velocity; see also Ref.~\cite{LLv2}.
    To derive Eq.~(\ref{Igen1})
from Eq.~(\ref{Igen}) we use the usual formulas of
transformations between the contrainvariant and coinvariant
vectors and the properties of the 
Killing, $t^\mu$ ($t^\mu\propto \omega$) and $\phi^\mu$,
 and normal, $\hat{n}_\nu$, vectors \cite{FIP86,NS03,MM26npa}.
 Therefore, the particle velocity
 $u_\mu$ can be expressed in terms of the Killing 
 vectors,
 $t^\mu$ and $\phi^\mu$, 
\be\l{utphi}
u_\mu=u^{t}\left(t_\mu +\omega \phi_\mu\right).
\ee
Then, one finds that
there is no contributions from the second term
(proportional to the pressure $\mathcal{P}$ and 
 gravitational metric $g^{}_{\mu\nu}$)
of the energy-momentum tensor, Eq.~(\ref{Tmn}),
and we
arrive at Eq.~(\ref{Igen1}).
Using also the metric elements from Eq.~(\ref{AWmetgen}), based on
Eqs.~(\ref{BLm}) and (\ref{SchwarzIN}),
for small rotational frequencies
$\omega$ one has
\be\l{Igen2}
I=\frac{1}{c^2}\int \epsi(n)e^{-\nu} r_\perp^2\left(\omega +
\frac{2c\tau}{r^2}\alpha^{}_K\right)\d \mathcal{V}~, 
\ee
where $\tau$ is given by Eq.~(\ref{14tauinexpsol}) for our gravitational metric
results.

Using approximately the asymptotic linear relation,
$\alpha^{}_K
\approx \omega \Theta/Mc$ [see Eq.~(\ref{OmegaK})],
in the second term of Eq.~(\ref{Igen2}) in the parentheses,
one finds 
\be\l{Theta2}
\Theta=\frac{\partial I}{\partial \omega}=
\frac{1}{c^2}\int \epsi(n)e^{-\nu} r_\perp^2\left(1 +
\frac{2\tau}{r^2}\frac{\Theta}{M}\right)\d \mathcal{V}~.
\ee
Solving this algebraic equation with respect to $\Theta$, with
the definition, Eq.~(\ref{rg}), for the gravitational
radius $r^{}_g$,
one obtains Eq.~(\ref{MI}).

\setcounter{equation}{0}
\renewcommand{\theequation}{C\arabic{equation}}

\section{Analytical asymptotic PSM outer solutions ($r>R$)}
\l{appC}

The asymptotic solutions of Eq.~(\ref{Xeqout}) for $\mathcal{X}(x)$ at
$x=r/r_g\gg 1$ with any accuracy can be found
analytically by using the standard power series method (PSM).
Let us look for the simplest solutions of this outer linear
homogenious equation of the second order, Eq.~(\ref{Xeqout}),
in the following PSM form of the power
series:
\be\l{eq17appF}
\mathcal{X}(x,q)=x^k(a_n x^n + a_{n-1}x^{n-1} + ...+ a_0 ).
\ee
Substituting this form to Eq.~(\ref{Xeqout}) and equaling the coefficients
in front of the each power of $x$ in the obtained identity, one finds at
the largest power $x^4$ the characteristic quadratic equation for $k$.
For example,
at $n=1$,
\be\l{eq18appF}
\mathcal{X}(x,q)=x^k(a^{}_1 x + a^{}_2),
\ee
the characteristic  equation writes
\be\l{eq19appF}
k^2+k-2 +q=0~.
\ee
The solutions of this quadratic equation is given by
\be\l{eq20appF}
k_{\pm}=\frac{-1\pm\sqrt{9-4q}}{2}~,
\ee
identically for any $a^{}_1$ and $a^{}_2$. For the convergent power series
    (\ref{eq18appF})
one has to take the root $k_{-}$. Then, at $x^3$ one obtains the $a^{}_2$
in terms of $a^{}_1$,
\be\l{eq21appF}
a^{}_2=\frac{a^{}_1 q}{4}~\frac{1-\sqrt{9-4q}}{2-q}.
\ee
Notice that the power series (\ref{eq18appF}) with $k=k_{-}$,
    Eq.~(\ref{eq20appF}), and $a^{}_2$ from Eq.~(\ref{eq21appF}) for
    any $a^{}_1$
    is identical to the Mathematica asymptotic solution,
    Eq.~(\ref{Xout2}) with $C_1=a^{}_1$.
For higher expansion powers $n$, Eq.~(\ref{eq17appF}), it is convenient
to use
Mathematica.
As seen from the characteristic equation (\ref{eq19appF})
these asymptotic solutions converge to zero at $x\gg 1$
for the $k_{-}$
root 
under the condition $q<9/4$~. 
Finally, we arrive to the
simplest asymptote
(\ref{Xout1}).

  \setcounter{equation}{0}
\renewcommand{\theequation}{D\arabic{equation}}

\section{To calculations of the surface mass,
      energy and
  MI components}
\label{appD}

Let us consider the integral,

\vspace{-0.5cm}
\be\l{Uint}
\mathcal{I}=\int_0^\infty \d r J(r) q(r) 
\left(
\frac{\partial \rho}{\partial r}\right)^2~,
\ee
where $J(r)$ is the radial Jacobian,
 Eqs.~(\ref{Jin}) and (\ref{Jout}), and
 $q(r)$ is another assumed
 smooth function of the radial coordinate $r$
as compared to the derivative $\partial \rho/\partial r$.
The 
latter is 
a bell-like function near the ES
 with a sharp maximum at $r=R$. 
 The 
function $q(r)$ is different 
 for surface components of the energy
$E_S$, mass $M_S$, and MI $\tilde{\Theta}_S$ and
 $\mathcal{T}_S$
in Eqs.~(\ref{enertot}), (\ref{MNStot}), and (\ref{adMIVS}),
respectively. 
Taking the assumed smooth functions $J(r)$ 
and $q(r)$ off the integral
at $r=R$, and
transforming the integration variable $r$ to $\rho$ by
$\rho=\rho(r)$, $d r=d \rho/(d \rho/dr)$, one can use 
the following equation \cite{MM24,MM26npa} for the derivative $\partial \rho/\partial r$,
\be\l{eq0}
\frac{\d \rho}{\d r}=-\sqrt{\frac{ \eps^{}_G(\rho)}{\mathcal{C}}}~,
\ee
where
$\eps^{}_G(\rho)$ is given by 
Eq.~(\ref{epsKG}).
As the approximate result
 up to a constant, one has
\be\l{Uint1}
\mathcal{I} \propto J(R) q(R) \frac{a}{R}\int_0^{\overline{\rho}} \d \rho 
\sqrt{ 
\epsilon^{}_G}~, 
\ee
where $\overline{\rho}$ is the inner  saturation density.
This integral $\mathcal{I}$ 
can be expressed 
in terms of the
tension coefficient
$\sigma \propto a/R$, Eqs.~(\ref{enertots}) and (\ref{leppar}).
Finally, we arrive at the expressions
for the  surface energy $E_S$, Eq.~(\ref{enertots}), 
    the mass $M_S$, Eq.~(\ref{MNSSfin}), and 
    the MI components $\tilde{\Theta}_S$,
   Eqs.~(\ref{adMItS}),  
and
$\mathcal{T}_S$, 
Eq.~(\ref{adMItfS}).
%
%
%
%

\bibliography{references}

\vspace{-1.5cm}
\begin{widetext}
  
\bigskip
\begin{center}
    \selectlanguage{ukrainian}
    
    {\bfseries\large МАКРОСКОПІЧНІ ОБМЕЖЕННЯ НА РАДІУС НЕЙТРОННОЇ ЗОРІ,
      ЗУМОВЛЕНІ ОБЕРТАННЯМ}
\end{center}
    \medskip

\vspace{-0.2cm}
    \begin{minipage}{0.99\linewidth}
      \begin{center}
       \small
        \noindent
     О.Г.~Магнер, А.А.~Улеєв, С.М.~Федоткін, У.В.~Григор$^\prime$єв\\

        Інститут ядерних досліджень Національної Академії Наук України,\\
        відділ теорії ядра, проспект Науки 47, Київ 03028, Україна,\\

        С.П.~Майданюк\\
        Інститут ядерних досліджень Національної Академії Наук України,\\
        відділ ядерних процесів,  проспект Науки 47, Київ 03028, Україна,\\

        Інститут сучасної фізики Китайської Академії наук\\
        Південний центр теорії ядра, Ксінкіао північна дорога 1,\\
        Хьюіжоу 516000, провінція Гуангдонг, Китай,\\
    
        А.~Бонасера, Т.~Депастас\\
        Циклотронний інститут Техасського A\&M університету,\\
        Льюдеке  Будинок 120, вул. Cпенса, Коледж Cтейшн, TX 77843, США\\

        Х.~Женг\\
        Школа фізики та інформаційної технології, Шаанксі Нормальний
        університет,\\
        Західна Чанжьян вул. 620,Чанжьян Дистрикт,
        Ксіан  710119, провінція Шаанксі, Китай,\\

        А.А.~Левон,\\
        Інститут ядерних досліджень Національної Академії Наук України,\\
        відділ фізики важких іонів, проспект Науки 47, Київ 03028, Україна,\\
        \end{center}
  \end{minipage}   

     \medskip
    
    \begin{minipage}{0.99\linewidth}
        \small
        \noindent{\bfseries Анотація.}
        Макроскопічну модель нейтронної зорі (НЗ) як скінченої ідеальної
        рідини в
        стані рівноваги узагальнено на системи, що обертаються. Це реалізовано
        завдяки лінійному збуренню за малою частотою
        поблизу зовнішньої-внутрішньої гравітаційної метрики Шварцшильда в межах
        підходу ефективної поверхні (ЕП). Кутовий момент
        та момент інерції (МІ) НЗ для повільного стаціонарного азимутального
        обертання
        навколо осі симетрії розраховано за допомогою метрики Керра у
        сферичних координатах 
        Боєра-Ліндквіста для зовнішньої та Хогана для внутрішньої метрик.
        Об’ємний та
        градієнтно-поверхневий члени макроскопічної густини енергії НЗ
        (рівняння стану) враховано в головному порядку лептодермічного
        параметра $a/R$, де $a$
— товщина кори ЕП, а $R$
        — ефективний радіус НЗ. Аналітичні вирази для макроскопічного МІ НЗ,
        $\Theta=\d I/\d \omega=\tilde{\Theta}/(1-\mathcal{T}_{t\varphi})$,
        отримано через статистично усереднений МІ, $\tilde{\Theta}$,
        та його кореляцію між часовою та азимутально кутовою змінними,
        $\mathcal{T}_{t\varphi}$,
у вигляді сум об’ємної та поверхневої компонент. МІ
суттєво змінюється залежно від ефективного радіуса $R$
через сильну гравітацію. Виявлено додаткове обмеження для радіуса НЗ у бік
менших доступних діапазонів, що зумовлено переважно
кореляціями та поверхневими внесками. Умови адіабатичності для застосовності
лінійної теорії збурень добре виконуються для багатьох НЗ із сильною
гравітацією та відносно великими періодами обертання.

\vspace{0.1cm}{\it Ключові слова:} нейтронні зорі, обертання, поверхневі деформації, гравітаційна
метрика, розширений підхід Томаса-Фермі, моменти інерції
    \end{minipage}
\end{widetext}
    

\end{document}